\documentclass[journal,10pt,doublecolumn]{IEEEtran}
\usepackage{amsmath,amsthm,amsfonts,amssymb,bm,mathrsfs}
\usepackage{extarrows}
\usepackage{graphicx,subfigure}
\usepackage{cite}
\usepackage{url}
\usepackage{algorithm}
\usepackage{algorithmic}
\usepackage{multirow,booktabs,threeparttable}
\usepackage{color}

\newcommand\dif{\mathrm{d}}

\allowdisplaybreaks

\theoremstyle{remark}
\newtheorem{lemma}{Lemma}
\newtheorem{theorem}{Theorem}
\newtheorem{corollary}{Corollary}

 \allowdisplaybreaks[4]
\begin{document}
\title{The Stochastic Geometry Analyses of Cellular Networks with $\alpha$-Stable Self-Similarity}


\author{\IEEEauthorblockN{Rongpeng Li,  Zhifeng Zhao, Yi Zhong, Chen Qi, and Honggang Zhang}\\
\thanks{R. Li, Z. Zhao, C. Qi and H. Zhang are with Zhejiang University, Hangzhou 310027, China (email:\{lirongpeng, zhaozf, qichen7c, honggangzhang\}@zju.edu.cn).}
\thanks{Y. Zhong is with the School of Electronic Information and Communications in Huazhong University of Science and Technology, Wuhan 430073, China (email: yzhong@hust.edu.cn).}
\thanks{This work was supported in part by National Key R\&D Program of China (No. 2018YFB0803702), National Natural Science Foundation of China (No. 61701439, 61731002), Zhejiang Key Research and Development Plan (No. 2018C03056).}
}
\maketitle
\begin{abstract}
To understand the spatial deployment of base stations (BSs) is the first step to analyze the performance of cellular networks and further design efficient networking protocols. Poisson point process (PPP), which has been widely adopted to characterize the deployment of BSs and established the reputation to give tractable results in the stochastic geometry analyses, usually assumes a static BS deployment density in homogeneous PPP (HPPP) models or delicately designed location-dependent density functions in in-homogeneous PPP (IPPP) models. However, the simultaneous existence of attractiveness and repulsiveness among BSs practically deployed in a large-scale area defies such an assumption, and the $\alpha$-stable distribution, one kind of heavy-tailed distributions, has recently demonstrated superior accuracy to statistically model the varying BS density in different areas. In this paper, we start with these new findings and investigate the intrinsic feature (i.e., the spatial self-similarity) embedded in the BSs. Afterwards, we refer to a generalized PPP setup with $\alpha$-stable distributed density and theoretically derive the related coverage probability. In particular, we give an upper bound of the derived coverage probability for high signal-to-interference-plus-noise ratio (SINR) thresholds and show the monotonically decreasing property of this bound with respect to the variance of BS density. Besides, we prove that our model could reduce to the single-tier HPPP for some special cases, and demonstrate the superior accuracy of the $\alpha$-stable model to approach the real environment.
\end{abstract}

\begin{IEEEkeywords}
Stochastic geometry, coverage probability, cellular networks, homogeneous Poisson point process (HPPP), $\alpha$-stable distributions, self-similarity
\end{IEEEkeywords}

\section{Introduction}
As a key enabler in the information and communications technology (ICT) industry, cellular networks play a decisive role in delivering communication messages and entertainment content \cite{wang_cellular_2014}. In order to meet the increasing traffic demand, cellular network operators gradually deploy different types of necessary infrastructure including lots of base stations (BSs). To understand the spatial deployment of BSs is the first step to facilitate the performance analyses of cellular networks and design efficient networking protocols. In the earliest stages, a two-dimensional hexagonal grid model was used, implying that BSs were deployed regularly, which deviates from the real scenarios. In the recent stages, Poisson point process (PPP) \cite{baccelli_stochastic_2010,baccelli_stochastic_2010-1,haenggi_stochastic_2012,andrews_tractable_2011} was assumed, which could roughly model the randomness in the realistic deployment of BSs in cellular networks, meanwhile leading to tractable results. However, given the total randomness assumption in PPP, its practical accuracy has been recently questioned \cite{guo_spatial_2013,lee_stochastic_2013}. Consequently, in order to reduce the modeling gap between the PPP model (especially the homogeneous PPP (HPPP) model) and the realistic spatial deployment of BSs, in-homogeneous PPP (IPPP) models, where the spatial density is location-dependent, have been proposed to measure the BS deployment in a large-scale area. Also, some efforts like inhomogeneous double thinning \cite{di_renzo_inhomogeneous_2018} have been put to tackle the non-stationary issue in IPPP.

On the other hand, in order to meet the larger traffic demand in certain areas (e.g., central business districts) in real life, BSs tend to be more densely deployed in these regions. Interestingly, according to an observation named ``preferential attachment" \cite{barabasi_emergence_1999}, Barab\'asi \textit{et al.} argued that many large networks should grow to be heavy-tailed. As its names implies, a heavy-tailed distribution has non-exponential bounded tail. Mathematically, for a heavy-tailed random variable $X$, the probability $\mathrm{Pr}(X>x)$ satisfies $\lim\limits_{x\rightarrow\infty} e^{\kappa x}\mathrm{Pr}(X>x) = \infty$, for all $\kappa >0$. Many well-known statistical distributions including power-law distribution (also named as generalized Pareto distribution), Weibull distribution, log-normal distribution, and $\alpha$-stable distribution \cite{samorodnitsky_stable_1994,li_learning_2017,zhou_alpha--stable_2015} belong to the heavy-tailed family. Therefore, heavy-tailed distributions appear to be more suitable to capture the societal feature for the practical BSs. 

Based on the practical BS deployment information, we have witnessed that the $\alpha$-stable distributions demonstrate superior accuracy to statistically model the large-scale BS deployments with the simultaneous existence of attractiveness and repulsiveness \cite{lee_stochastic_2013,chun_modeling_2015,guo_spatial_2013,miyoshi_spatial_2016, chiu_stochastic_2013} and also outperform other aforementioned distributions to characterize the spatial BS deployment density \cite{zhou_alpha--stable_2015,zhou_large-scale_2015,chiaraviglio_what_2016,chiaraviglio_reality_2016}. But the density alone cannot reach some intuitive conclusions. Fortunately, the $\alpha$-stable distributions, which have been widely adopted to characterize the distribution of aggregated traffic at the BS level in cellular networks and at the switch level in wired broadband networks, often imply the self-similarity of traffic \cite{li_understanding_2015,li_learning_2017}. As a popular property in complex networks \cite{song_self-similarity_2005}, the self-similarity of traffic in the temporal domain means that the distribution of traffic remains invariant under different temporal scales. Similar to the case of traffic, as the practically deployed BSs are $\alpha$-stable distributed, it is natural to ask whether the spatial self-similarity holds in cellular networks. If yes, compared to other point process models, a generalized PPP setup with $\alpha$-stable distributed density (or the $\alpha$-stable model for short) sounds to become applicable with significant merit, since the $\alpha$-stable model facilitates to explain the density differences in practical BS deployments which conventional HPPP is not qualified for, but still avoids the delicate design of location-dependent density function for IPPP models.
\subsection{Related Works}
\label{sec:related}
As discussed later in Section \ref{sec:survey}, the PPP model has provided many useful performance trends. However, the concern about total independence between nodes (e.g., BSs) has never stopped. Hence, in order to reduce the modeling gap between the single-tier HPPP model and the practical BS deployment, some researchers have adopted two-tier or multiple tier HPPP models \cite{elsawy_two-tier_2014,cheung_throughput_2012,zeinalpour-yazdi_outage_2014,soh_cognitive_2013,cao_improving_2013,cao_optimal_2013,dhillon_modeling_2012,renzo_stochastic_2015}. Though these models may lead to some tractable results, but it lacks the reasonable explanation to divide the gradually deployed and increasingly denser heterogeneous cells \cite{ge_5g_2016} into different tiers. On the contrary, the $\alpha$-stable model contributes to understanding the spatial self-similarity and bridging the gap between cellular networks and other social behavior-based complex networks.

On the other hand, point processes with either attractive or repulsive spatial correlations have been explored as well. 
\begin{itemize}
	\item For attractive point processes, \cite{lee_stochastic_2013} has shown that Poisson cluster process is more suitable to model the spatial distribution of BSs in urban areas. The general clustering nature of deployed BSs in highly populated urban areas clearly reflects the aggregation property of ever-growing traffic demands in cellular networks \cite{zhou_large-scale_2015}. \cite{chun_modeling_2015} has further verified the aggregated interference when the transmitting nodes are modeled by Poisson cluster process and compared the corresponding results with that in the HPPP model. 
	\item On the contrary, \cite{guo_spatial_2013} and \cite{miyoshi_spatial_2016} have argued that BSs, in particular macro BSs, in cellular networks tend to be deployed systematically, such that any two BSs are not too close. Thus, a spatial model based on a point process with the repulsive nature seems to be more desirable \cite{guo_spatial_2013,miyoshi_spatial_2016}. Furthermore, the Mat\'ern hard core point process (MHCP) model \cite{chiu_stochastic_2013} possesses limited tractability and leaves many open challenges to be addressed. The Ginibre point process (GPP) model, one typical example of determinantal point processes on the complex plane, has been widely adopted and shows a promising trade-off between accuracy and tractability  \cite{deng_ginibre_2015,miyoshi_cellular_2014,kobayashi_uplink_2014}. In particular, Deng \textit{et al.} \cite{deng_ginibre_2015} have shown that the GPP leads to the same trend curve of coverage probability\footnote{The coverage probability indicates the probability that the signal-to-interference-plus-noise ratio (SINR) for a UE achieves a target threshold.} as the HPPP. 
\end{itemize}

Hence, we can come to the following conclusion that the spatial BS density of large-scale cellular networks simultaneously possesses two conflicting features, that is, the density is very large in some clustering regions while being significantly smaller in others. The conclusion is also consistent with our common sense. Furthermore, the heavy-tailed distributions, which generate small values with the high probability but still allow comparatively larger values, fit well to model the spatial BS density. Our previous works \cite{zhou_large-scale_2015,zhou_alpha--stable_2015,chiaraviglio_what_2016,chiaraviglio_reality_2016} have shown that the $\alpha$-stable distribution, one kind of heavy-tailed distributions, could be used to more accurately model the spatial BS density in China and Italy, especially in urban areas. $\alpha$-stable distributions also owe their importance in both theory and practice to the generalization of the central limit theorem \cite{samorodnitsky_stable_1994} and the accompanying self-similarity \cite{crovella_self-similarity_1997,qi_characterizing_2016} of the stable family. Hence, as the spatial BS density could be better modeled by the $\alpha$-stable distributions, it is essential to theoretically examine the spatial self-similarity and understand its impact on coverage probability. In other words, we will take the very first step to investigate the spatial self-similarity in the BS deployment and perform stochastic geometry analyses of cellular networks when the BS density is $\alpha$-stable distributed.

\subsection{Contributions}

Different from the previous works \cite{zhou_alpha--stable_2015,zhou_large-scale_2015,chiaraviglio_reality_2016,chiaraviglio_what_2016}, where we mainly validate the universal superior accuracy of $\alpha$-stable distribution to model the practical base station (BS) deployment density in cellular networks, this paper aims at leveraging the $\alpha$-stable model to theoretically study the coverage probability of cellular networks in a large-scale area. Belonging to one of the precursor works, we take advantage of a large amount of realistic BS deployment records and provide the following key insights:

\begin{itemize}
	\item Firstly, on top of the $\alpha$-stable model validated in  \cite{zhou_alpha--stable_2015,zhou_large-scale_2015,chiaraviglio_reality_2016,chiaraviglio_what_2016} this paper mathematically studies the coverage probability in a large-scale area and gives an upper bound for high SINR thresholds. We talk about the monotonicity of this bound with respect to the variance of BS density. This paper also theoretically proves that for some special cases, our result could be reduced to that for the HPPP model in \cite{andrews_tractable_2011}.

	\item Secondly, during the mathematical derivation, this paper leverages the self-similarity in the spatial deployment of BSs, thus laying the foundation to apply the advance in complex network theory to examine cellular networks. Particularly, the self-similarity is verified based on the practical BS deployment records from both cellular network operators and the open database.
	
	\item Thirdly, this paper compares the coverage probability between the models and the real BS deployment in both Hangzhou (China) and Rome (Italy) and shows the superior accuracy of the $\alpha$-stable model. Besides, this paper studies the coverage probability performance under extensive simulation settings and validates that the simulation results are consistent with our theoretical derivations.
\end{itemize}

The remainder of the paper is organized as follows. Section \ref{sec:survey} provides a brief application survey of stochastic geometry. In Section \ref{sec:background}, we present some necessary mathematical background, introduce the realistic dataset of spatial BS deployment and validate the spatial self-similarity. In Section \ref{sec:analyses}, we provide the analyses of coverage probability. Section \ref{sec:performance} evaluates the computable representation obtained by the theoretical analysis and compares it with both the real environment and the single-tier HPPP model. Finally, we conclude this paper in Section \ref{sec:conclusion}.

\section{Brief Application Survey of Stochastic Geometry}
\label{sec:survey}
Stochastic geometry tools have successfully established the reputation to model and analyze wireless networks, as they are able to capture the topological randomness in the network geometry and lead to tractable analytical results \cite{andrews_femtocells:_2012,andrews_overview_2013,haenggi_stochastic_2009,andrews_tractable_2011,elsawy_modeling_2017,flint_performance_2015,elsawy_two-tier_2014,haenggi_interference_2009,chun_modeling_2015,guo_gauss-poisson_2016, chiu_stochastic_2013,miyoshi_cellular_2014,kobayashi_uplink_2014,li_fitting_2014,renzo_intensity_2016}. There is no doubt that the PPP belongs to the most popular point process used in the literature because of its tractability and has led to a lot of meaningful research results for cellular networks with different kinds of cutting-edge techniques. The baseline operations in single-tier and/or multiple-tier downlink cases are examined in \cite{andrews_tractable_2011,guo_spatial_2013,cheung_throughput_2012,mukherjee_distribution_2012,yu_dynamic_2015,dhillon_modeling_2012,blaszczyszyn_using_2013,renzo_average_2013,renzo_stochastic_2015,lu_stochastic_2015,renzo_stochastic_2014} (and references therein) while their counterparts in uplink cases are considered in references like \cite{renzo_stochastic_2016,martin-vega_analytical_2016,novlan_analytical_2013,singh_joint_2015,lee_uplink_2014,zeinalpour-yazdi_outage_2014}. User association and load balancing are examined in \cite{mirahsan_hethetnets:_2015,lin_optimizing_2015,dhillon_load-aware_2013,singh_offloading_2013}. Cognitive, cooperative and intelligent cellular networks are taken into account in \cite{elsawy_two-tier_2014,elsawy_hetnets_2013,soh_cognitive_2013,huang_analytical_2013,lima_statistical_2013,semasinghe_evolutionary_2015}. Energy efficiency, energy harvesting, and BS sleeping for green cellular networks are investigated in \cite{deng_modeling_2016,li_throughput_2014,cao_optimal_2013,cao_improving_2013,soh_energy_2013,renzo_system-level_2017,renzo_system-level_2018}. Physical layer security is examined in \cite{wang_uncoordinated_2015,chen_secrecy_2017,pinto_secure_2012,wang_opportunistic_2016, yang_delivery-secrecy_2018}.
Besides, multiple-input multiple-output (MIMO) antenna, in-band full-duplex (FD), and millimeter wave systems are studied in
\cite{renzo_mathematical_2014,lee_spectral_2015,gupta_downlink_2014,afify_unified_2016,bai_analyzing_2016,singh_tractable_2015,muhammad_cell_2017,atzeni_full-duplex_2015,lee_hybrid_2015,tabassum_analysis_2016,bai_coverage_2015,renzo_stochastic_2015,renzo_stochastic_2015-1,shojaeifard_massive_2017,lam_system-level_2016,shojaeifard_self-interference_2017}. Also, the PPP-related works have been extended to mobile ad hoc networks \cite{huang_spatial_2012}. Along with the continuous progress in techniques for physical-layer and network management in wireless networks, the PPP model offers a useful theoretical platform for the initial performance calibration, so as to save economic expenditure and avoid time-consuming large-scale setup for realistic tests. For example, FD communication, which benefits from the advance in signal processing and self-interference cancellation techniques, is optimistically promoted to double the spectral efficiency for wireless networks \cite{li_full-duplex_2017}. However, given the more complicated interference in FD networks, it is essential to conduct a careful re-examination. Consistent with the system-level simulation results in \cite{li_full-duplex_2017}, \cite{alammouri_-band_2016} has taken advantage of the PPP model and demonstrated the potential negative effect that FD communications might impose on the uplink transmission.

\section{Mathematical Background and Statistical Modeling}
\label{sec:background}

\subsection{Mathematical Background}
Beforehand, Table \ref{tb:notations} summarizes the most used notations in this paper. 

\begin{table*}
	\centering
	\caption{A list of the main symbols and functions in the paper.}
	\label{tb:notations}
	\begin{tabular}{c | l | l}
		\toprule
		Symbol & Meaning & Default Value in Sec. \ref{sec:performance} \\
		\midrule
		$p_c$ & The coverage probability & --\\
		\hline
		$p_d(r)$ & The PDF of the distance $r$   & --\\ 
		\hline
		$\mathbb{S}(\alpha,\sigma, \mu)$ & $\alpha$-stable distributions with stability $\alpha$, skewness $\beta=1$, scale $\sigma$ and shift $\mu$ & $\alpha = 0.6$, $\sigma = 0.25$ and $\mu = 0.25$   \\ 
		\hline 
		$\delta$ & Pathloss exponent & $\delta = 4$\\
		\hline
		$h$ & The Rayleigh fading gain for the signal link & -- \\
		\hline
		$g$ & The fading exponent for the interfering links &  Assumed to follow Rayleigh fading.\\
		\hline
		$\zeta$ & Rayleigh fading factor & $\zeta = 1$\\ 
		\hline
		$N_0$ & Noise factor & $N_0 = 1$ (i.e., SNR = $0$ dB)\\
		\hline
		$H$ & The Hurst parameter & $H = 0.9$\\
		\hline
		$a$ & The self-similarity zooming parameter & $a = 2$\\
		\hline 
		$R$ & The predefined radius of inner circle in Fig. \ref{fig:selfSimilarityIllustration} & $R = 40$\\
		\hline 
		$\lambda$ & The spatial density of BSs & --\\
		\hline
		$N(r)$ & The number of BSs within an $r$-radius circle & --\\
		\hline
		$\Gamma(d)$ & The standard Gamma function $\Gamma(d)=\int_{0}^{\infty} t^{d-1} e^{-t} \dif t$ for $d > 0$ & --\\
		\hline
		$\Gamma(d,x)$ & The incomplete Gamma function $\Gamma(d,x)=\int_{x}^{\infty} t^{d-1} e^{-t} \dif t$ for $d > 0$  & -- \\
		\hline
		$\Theta(s,b,c)$ & $\Theta(s,b,c) \stackrel{\textrm{def}}{=}  (s g)^{2/\delta} \left(\Gamma(-\frac{2}{\delta}+1, s g b^{-\delta}) - \Gamma(-\frac{2}{\delta} +1, s g c^{-\delta}) \right)  $ & --\\
		\hline
		$\Lambda(s,b,c)$ & $\Lambda(s,b,c) \stackrel{\textrm{def}}{=}    c^2 \left[ 1 - \exp \left( -s g c^{-\delta} \right) \right] - b^2 \left[ 1 - \exp \left( -s g b^{-\delta} \right) \right]  $ & --\\
		\hline
		$\Xi(s,b,c)$ & $\Xi(s,b,c) \stackrel{\textrm{def}}{=}  \pi \mathbb{E}_g \Big( \Lambda(s,b,c) - \Theta(s,b,c)\Big) $ & --\\
 		\bottomrule
	\end{tabular}
\end{table*}

\subsubsection{$\alpha$-Stable distributions}
An $\alpha$-stable distributed random variable does not always possess a closed-form probability density function (PDF). Instead, $x$ is often defined by its characteristic function. Specifically, $x$ is said to obey the $\alpha$-stable distribution $f(x)$ if there are parameters $0<\alpha \leq 2$, $\sigma \geq 0$, $-1\leq \beta \leq 1$, and $\mu \in \mathbb{R}$ such that it has a characteristic function with the form:
\begin{align}
\label{eq:characteristic_func_stable}
&\Phi(\omega)= \mathbb{E}\left(\exp (j\omega x)\right)\\
=&\left\{
\begin{aligned}
&\exp\left\{-\sigma^{\alpha} \vert\omega\vert^{\alpha} \left(1-j\beta \text{sgn} (\omega) \tan \left(\frac{\pi \alpha}{2} \right) \right) + j\mu \omega \right\}, \alpha\neq 1;\\
&\exp\left\{-\sigma \vert\omega\vert \left(1+j\frac{2\beta}{\pi} \text{sgn} (\omega) \ln\vert\omega\vert \right) + j\mu \omega \right\}, \alpha= 1.\\
\end{aligned}
\right.\nonumber
\end{align}
The function $\mathbb{E}(\cdot)$ represents the expectation operation with respect to a random variable. $\alpha$ is called the characteristic exponent and indicates the index of stability, while $\beta$ is identified as the skewness parameter. $\alpha$ and $\beta$ together determine the shape of the distribution. Moreover, $\sigma$ and $\mu$ are called scale and location shift parameters, respectively. Another direct definition for a stable distribution is that a linear combination of two independent identically distributed random variables has the same distribution \cite{samorodnitsky_stable_1994}. In particular, if $\alpha=2$, the $\alpha$-stable distribution reduce to the Gaussian distribution. Since the characteristic function can be used to derive moments of a random variable, leveraging the $\alpha$-stable distribution to model the BS density brings many statistical merits explicitly.

Our previous works \cite{zhou_alpha--stable_2015,chiaraviglio_what_2016} have validated that $\beta=1$ holds for the fitting results of actual spatial BS density in both Hangzhou, China and Rome, Italy. For simplicity of representation, we use the operator $b \sim c$ to denote that $b$ and $c$ have the same distribution and further have $\lambda \sim \mathbb{S}(\alpha,\sigma, \mu)$ to indicate that the spatial BS density $\lambda$ follows the $\alpha$-stable distributions with $\beta=1$. Moreover, if a random variable $x \sim \mathbb{S}(\alpha,\sigma, 0)$, $x$ has a corresponding Laplace transform (Proposition 1.2.12, \cite{samorodnitsky_stable_1994}) as
\begin{equation}
\Psi(s) = \mathbb{E} \left(\exp (-s x)\right)	=\left\{
\begin{aligned}
& \exp\left\{ -\frac{\sigma ^{\alpha}}{\cos \frac{\pi \alpha}{2}} s^{\alpha} \right\}, \alpha\neq 1;\\
&\exp\left\{ \sigma \frac{2}{\pi} s \ln s \right\}, \alpha= 1.\\
\end{aligned}
\right.
\label{eq:stableLap}
\end{equation}

\subsubsection{The self-similarity}
\label{sec:self-similarity}
The self-similarity has been mostly applied to model time series \cite{crovella_self-similarity_1997}. Given zero-mean, stationary time series $Y=(Y_t, t=1, 2, 3, \cdots)$, $Y$ is called as self-similar process if its $m$-aggregated series $Y^{(m)}=(Y^{(m)}_t, t=1, 2, 3, \cdots)$ with each element $Y^{(m)}_t = \sum\limits_{k=(t-1)m+1}^{tm} Y_k$ satisfy $Y^{(m)}_t \sim m^H Y_t$ for all $m>0$, where $H \in (0,1)$ is the Hurst parameter indicating the decay rate of statistical dependence of two points with increasing time interval. In other words, the self-similarity implies that the time series is exactly or approximately similar to a part of itself.

The concept of self-similarity is also applicable to spatial scenarios. For example, H. ElSawy \textit{et al.}  \cite{elsawy_modeling_2017} have shown the aggregated interference from distributed nodes to a receiver obey $\alpha$-stable distributions, while X. Ge \textit{et al.} \cite{ge_wireless_2016} have verified that the statistical characteristics of the wireless cellular coverage boundary possess the self-similarity or fractal property.

However, no exact self-similar phenomenon exists in the real world. Most of the self-similar phenomena observed in the real world only approximately have the statistical characteristic. Moreover, the self-similarity of random processes is usually evaluated by the Hurst parameter, which can be estimated using two typical methods (i.e., the rescaled adjusted range statistic (R/S) method and the variance-time (V-T) method \cite{ge_wireless_2016}). Usually, the hurst parameter $H$ and a larger $H \in (0.5,1)$ corresponds to stronger self-similarity.

\subsection{Verification of Self-Similarity in BSs}

\begin{figure*}
	\centering
	\includegraphics[width=0.875\textwidth]{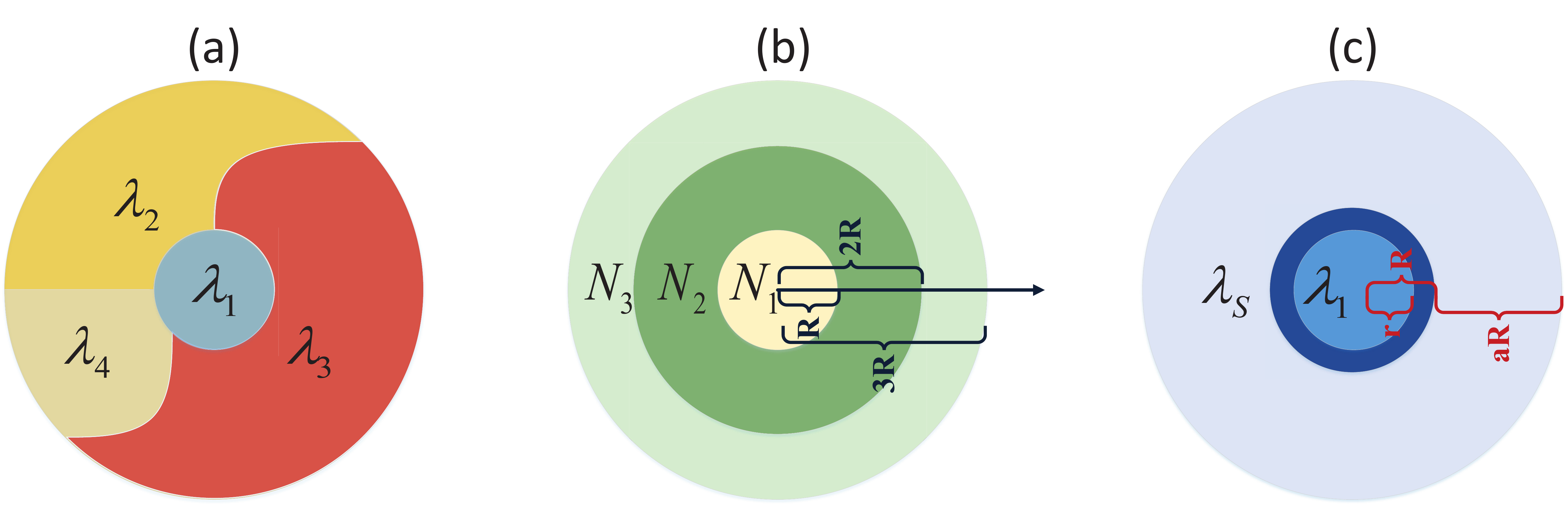}
	\caption{An illustration of the self-similarity for the spatial BS deployment.}
	\label{fig:selfSimilarityIllustration}
\end{figure*}

As validated in \cite{zhou_alpha--stable_2015,chiaraviglio_what_2016}, when we divide the region into several parts and calculate the corresponding PDF of the spatial BS density in each part, we observe that the PDF follows the $\alpha$-stable distributions. For example, Fig. \ref{fig:selfSimilarityIllustration}(a) depicts that a region is divided into four parts. The spatial BS density in each part $\lambda_1$, $\lambda_2$, $\lambda_3$ and $\lambda_4$ are mutually dependent and obey the $\alpha$-stable distributions with the same values $\alpha$, $\sigma$ and $\mu$. The correlation between BS density in different parts makes it challenging to directly derive the coverage probability. Instead, it is essential to take the statistical modeling of the correlation into consideration. In that regards, self-similarity emerges as a promising technique to characterize the correlation of the spatial density, as it has manifested its importance and effectiveness by modeling the correlation in different scales. 

\begin{figure*}
	\centering
	\includegraphics[width=0.85\textwidth]{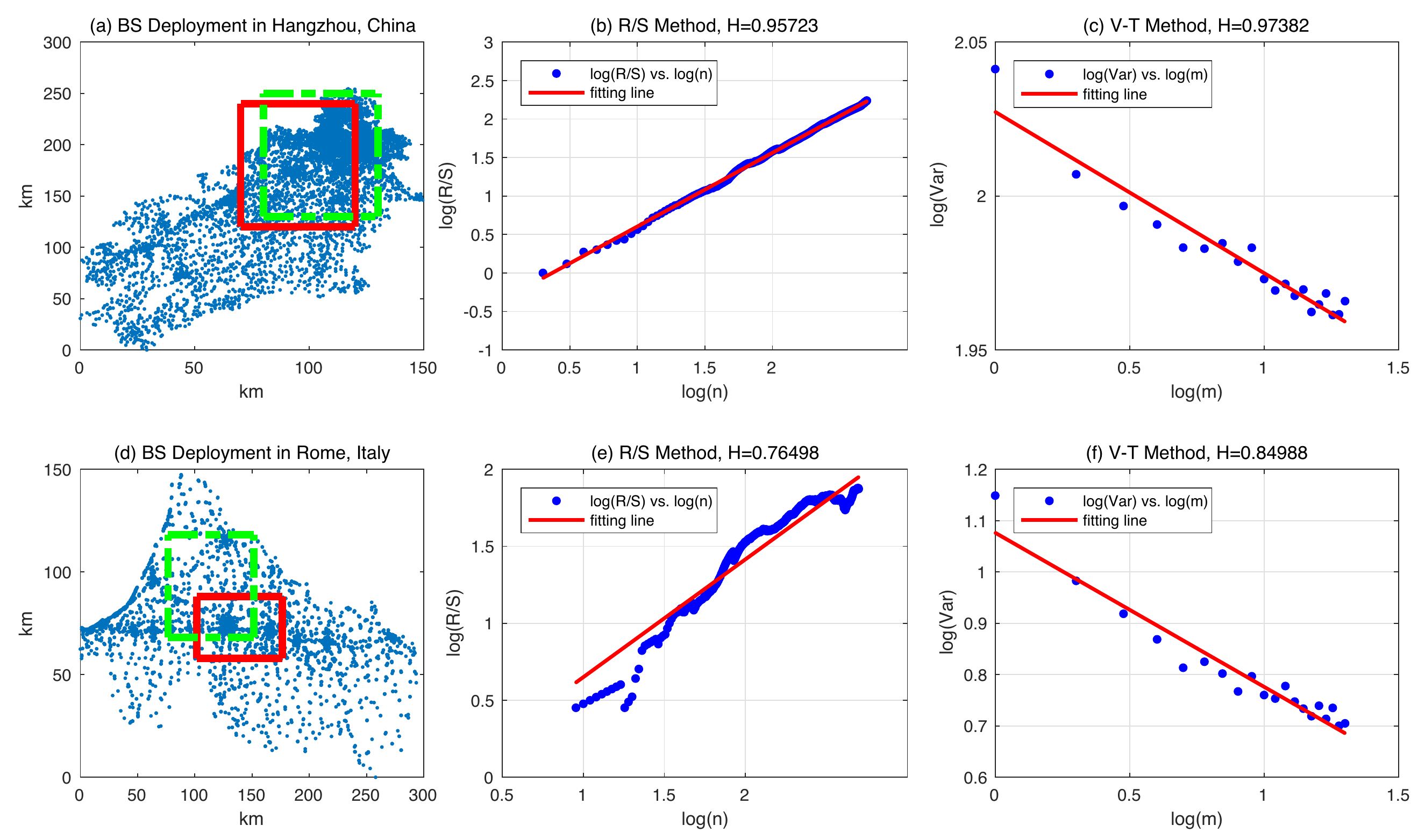}
	\vspace{-0.5cm}
	\caption{The self-similarity modeling results for the spatial BS deployment in Hangzhou and Rome.}
	\label{fig:self-similarity-val}
\end{figure*}

In order to show the self-similarity in spatial BS deployment, we take advantage of the collected BS deployment records in Hangzhou, China and Rome, Italy\footnote{Interested readers could visit \url{http://www.rongpeng.info/files/sup_file_stable.pdf} to find more results for the BS deployment in other four cities (i.e., Paris (France), Seoul (South Korea), Munich (Germany), Warsaw (Poland)), based on the open database from OpenCellID (\url{https://opencellid.org/}).}. Fig. \ref{fig:self-similarity-val}(a) and \ref{fig:self-similarity-val}(d) illustrate the corresponding deployment situation in both cities. Similar to the common validation process in temporal dimension  \cite{crovella_self-similarity_1997}, we verify the accuracy of self-similarity in spatial BS deployment as follows. As depicted in Fig. \ref{fig:selfSimilarityIllustration}(b), we randomly select a point as the starting ``origin point" and thus get some concentric circles with increasingly larger radii (e.g., from $R$ to $2R$, $3R$, $\cdots$). Then, we could get a series, each value corresponding to the number of BSs in one circle. Afterwards, we could apply the aforementioned R/S method and V-T method in Section \ref{sec:self-similarity}. Fig. \ref{fig:self-similarity-val}(b)(c) and Fig. \ref{fig:self-similarity-val}(e)(f) provide the corresponding log-log plots for both cities. From these figures, it can be observed that all the estimated Hurst parameters are very close to $1$. Therefore, we could boldly argue that the spatially deployed BSs possess the spatial self-similarity. Then, according to the definition of self-similarity, the number of BSs $N(a R)$  within a circle of radius $aR$ should satisfy that  $N(a R) \sim a^H N (R), \forall R, a>0, 0\leq H<1 $, where $a$ indicates the self-similarity zooming parameter and $H$ denotes the Hurst parameter.

\section{The Coverage Probability Analyses}
\label{sec:analyses}
\subsection{The System Model and Its Stationarity}
In this part, we derive the coverage probability in a downlink cellular network within a region of interest $\mathcal{R}$. Specifically, the coverage probability is defined as the probability that the SINR for a UE achieves a target threshold. Mathematically, the coverage probability is formulated as
\begin{equation}
	\mathbb{P}(\mathrm{SINR} > T) = \mathbb{P} \left[ \frac{hr^{-\delta}}{N_0 + I_r} > T \right],
\end{equation}
where $\delta$ denotes the pathloss exponent factor of the standard propagation channel. $N_0$ is the additive white Gaussian noise (AWGN) factor and $T$ is the target threshold. The fading is assumed to follow the Rayleigh fading, that is, $h \sim \exp(\zeta)$. 

For BSs deployed in the region $\mathcal{R}$, the $\alpha$-stable model assumes that BSs are characterized by a generalized PPP where for any infinitesimally small Borel set $\mathcal{B} \in \mathcal{R}$, the BS deployment density within $\mathcal{B}$ satisfies the $\alpha$-stable distribution (i.e., $\lambda \sim \mathbb{S}(\alpha,\sigma, \mu)$) and the corresponding number of BSs is 
$\mathbb{P}(N=n) = e^{-\lambda \Vert \mathcal{B} \Vert} \frac{(\lambda \Vert \mathcal{B} \Vert)^n}{n}$ where $\Vert \mathcal{B} \Vert$ is the size of the set $\mathcal{B}$. In other words, the $\alpha$-stable model is a doubly stochastic point process. According to \cite{dario_properties_2011}, a doubly stochastic point process is stationary if and only if the intensity is stationary. As the BS deployment density is assumed to follow the same distribution for any infinitesimally small Borel set $\mathcal{B} \in \mathcal{R}$ and thus the distribution of $\lambda$ is translation-invariant, the $\alpha$-stable model is stationary.

On the other hand, we also assume that users are spatially distributed according to a stationary point process independent of the BS deployment. Therefore, without loss of generality, we could further assume the UE is located at the ``origin point". When the distance from a UE to its serving BS $b_0$ is $r$, $I_r$ denotes the cumulative interference from all other BSs $i$ (except BS $b_0$) to the UE.

\subsection{Impact of $\alpha$-Stable Distributions and Self-similarity}
\label{sec:impact_ss}
Before delving into the coverage probability, we first leverage the $\alpha$-stable distributions and the self-similarity to further shape the BS deployment. As depicted in Fig. \ref{fig:selfSimilarityIllustration}(c), we assume that there exists a specific $R$ (i.e., $R > r$) to divide the whole region of interest into two parts. Besides, if the spatial BS density $\lambda$ for the region within the radius $R$ satisfies $\lambda \sim \mathbb{S}(\alpha,\sigma,\mu)$, the Laplace transform of $\lambda$ could be achieved by directly applying \eqref{eq:stableLap} and is given in the following lemma.

\begin{lemma}
\label{lemma:stableLap_lambda}
The spatial BS density $\lambda \sim \mathbb{S}(\alpha,\sigma,\mu)$ has a Laplace transform
\begin{equation}
\begin{aligned}
& \Psi(s) \\
=& \mathbb{E} \left(\exp (-s \lambda)\right)\\
=&\left\{
	\begin{aligned}
	& \exp\left\{ -\frac{\sigma ^{\alpha}}{\cos \frac{\pi \alpha}{2}} s^{\alpha} -\mu s \right\}, \alpha\neq 1;\\
	&\exp\left\{\frac{2  \sigma }{\pi} s \ln s  -\mu s \right\}, \alpha= 1.\\
	\end{aligned}
	\right.	
\end{aligned}
\label{eq:stableLap_lambda}
\end{equation}
\end{lemma}

Moreover, we can obtain the following lemma to characterize the spatial BS density $\lambda_S$ for the region outside the radius $R$ if $\lambda \sim \mathbb{S}(\alpha,\sigma,\mu)$.

\begin{lemma}
\label{lemma:laplace_lambda_s}
The spatial BS density $\lambda_S$ will follow $\alpha$-stable distributions (i.e., $\lambda_{S}  \sim \lambda a^{H-2}$) and the corresponding Laplace transform could be formulated as
\begin{equation}
\begin{aligned}
& \Psi_{S}(s)  \\
= & \mathbb{E} \left(\exp (-s \lambda_S)\right) \\
= & \left\{
	\begin{aligned}
	& \exp\left\{ -\frac{\sigma ^{\alpha} a^{\alpha (H-2)} }{\cos \frac{\pi \alpha}{2}} s^{\alpha} -s a^{H-2} \mu  \right\}, \alpha\neq 1;\\
	&\exp\left\{\frac{2 \sigma a^{H-2}}{\pi}  s \left[\ln s + (H-2) \ln a \right] -s a^{H-2} \mu \right\}, \alpha= 1.\\
	\end{aligned}
	\right.	
\end{aligned}
\end{equation}
\end{lemma}

\begin{proof}
Following the definition of self-similarity, the number of BSs in the interfering region is coupled with the self-similarity zooming parameter $a$ and could be formulated as $N(aR)=\lambda_{S} \pi a^2 R^2 \sim  a^H N(R) = \lambda \pi a^H R^2$. Therefore, 
\begin{equation}
\label{eq:lambda_self_similar}
\lambda_{S}  \sim \lambda a^{H-2}.
\end{equation}
	
Then, the Laplace transform of $\lambda_S$ could be derived as
\begin{equation}
\begin{aligned}
\Psi_{S}(s) & = \mathbb{E} \left(\exp (-s  a^{H-2} (x +\mu)) \right) \\
& = \Psi(s  a^{H-2})	\cdot \exp\left\{ -s a^{H-2} \mu \right\}.
\label{eq:selfSimilarityStableLapDer}
\end{aligned}
\end{equation}

By merging \eqref{eq:stableLap} and \eqref{eq:selfSimilarityStableLapDer}, we obtain the result.
\end{proof}

\subsection{The PDF of the Distance from Nearest BS to the user equipment}

In this paper, we adopt the minimum distance as the user association metric. Therefore, when the distance from a UE to its nearest BS (or the serving BS) $b_0$ is $r$, all distances from the interference BSs to the target user equipment (UE) must be larger than $r$. The derivation methodology of the PDF of $r$ could basically follow the well-established lines in \cite{andrews_tractable_2011}. However, as the BS density varies in space and obeys the $\alpha$-stable distributions, the derivations should be re-considered. Therefore, we give the following the theorem.
\begin{figure*}[!t]
	%\newcounter{mytempeqncnt}	
	\normalsize
	\begin{equation}
	\label{eq:pdf_r}
	p_d(r)  = \left\{
	\begin{aligned}
	& \exp\left\{ -\mu \pi r^2 -\frac{(\pi \sigma) ^{\alpha}}{\cos \frac{\pi \alpha}{2}} r^{2\alpha} \right\}  \left(2\mu \pi r  + 2\alpha \frac{(\pi \sigma) ^{\alpha}}{\cos \frac{\pi \alpha}{2}}  r^{2\alpha -1}  \right), \alpha\neq 1;\\
	&\exp\left\{  -\mu \pi r^2+2 \sigma r^2 \ln (\pi r^2)\right\} \left( 2\mu\pi r + 4\sigma r \ln (\pi r^2) + 4\sigma r \right), \alpha= 1.\\
	\end{aligned}
	\right.
	\end{equation}
	\hrulefill
	\vspace*{4pt}
\end{figure*}

\begin{theorem}
\label{theorem:pdf_r}
The PDF of the distance $r$ from the closest BS to the target UE in a cellular network with $\alpha$-stable distributed BS density is given in \eqref{eq:pdf_r} on Page \pageref{eq:pdf_r}.
\end{theorem}

This theorem could be easily obtained by applying Lemma \ref{lemma:laplace_lambda_s} to derive the probablity of no BS closer than $r$, as the lines in \cite{andrews_tractable_2011}.

\begin{corollary}
\label{cor:pdf_r}
For HPPP with static density $\lambda$, the PDF of $r$ could reduce to $p_d(r) = \exp(-\lambda \pi r^2) 2\pi \lambda r$. Similarly, for varying distribution $\lambda$, the conditional probability $\mathbb{P}(r|\lambda) = \exp(-\lambda \pi r^2) 2\pi \lambda r$.
\end{corollary}

This corollary could be achieved by directly applying Theorem \ref{theorem:pdf_r} with a static density $\lambda=\mu$ and thus $\sigma=0$. It is also consistent with the conclusions in \cite{andrews_tractable_2011}. Also for the $\alpha$-stable with small $\sigma$, the difference in $p_d(r)$ between the $\alpha$-stable model and the HPPP is quite small.

\subsection{Main Result}

We now state our main result for the coverage probability analysis. Generally, the coverage probability $p_c$ is mainly determined by the SINR threshold $T$, the spatial BS density distribution $\mathbb{S}(\alpha,\sigma,\mu)$, and the pathloss exponent $\delta$. In some sense, $p_c(T,\mathbb{S}(\alpha,\sigma,\mu),\delta)$ is tightly coupled with the cumulative interference $I_r$ for the $\alpha$-stable distributed and self-similar BS deployment. The following lemma characterizes the relationship between $p_c(T,\mathbb{S}(\alpha,\sigma,\mu),\delta)$ and $I_r$.

\begin{lemma}
	\label{lemma:coverage_prob}
	The coverage probability $p_c(T,\mathbb{S}(\alpha,\sigma,\mu),\delta)$ could be obtained from the following formula, that is,	
	\begin{equation}
		\begin{aligned}
		&p_c(T,\mathbb{S}(\alpha,\sigma,\mu),\delta) \\
		=&\iint_{r>0,\lambda>0} e^{-\zeta T r^{\delta} N_0 }\mathcal{L}_{I_r}( \zeta T r^{\delta}) \mathbb{P}(r|\lambda) f(\lambda) \dif r\dif \lambda,	
		\end{aligned}
	\end{equation}
	where $\mathcal{L}_{I_r}(s)$ is the Laplace transformation of random variable $I_r$ evaluated at $s$ conditioned on the distance to the closest BS from the origin. The Rayleigh fading coefficient is assumed to satisfy $h \sim \exp(\zeta)$.
\end{lemma}
Similar to \cite{andrews_tractable_2011}, Lemma \ref{lemma:coverage_prob} could be simply obtained by calculating the probability conditioning on the distance $r$ from the nearest serving BS to the UE.

Next, we focus on how to calculate the Laplace transform of the cumulative interference $\mathcal{L}_{I_r}(s)$, when the BS deployment obey $\alpha$-stable distributions and self-similarity. In Section \ref{sec:impact_ss}, we divide the region of interest into two concentric circles. For the inner circle $\Delta$ with a radius $R$, the impact of the spatial BS density $\lambda$ is similar to the impact of $r$ and explicitly represented by the latter's PDF $p_d(r)$. Meanwhile, the outer circle $\Delta_S$ with the radius spanning from $R$ to $aR$ has a spatial BS density $\lambda_S \sim \lambda a^{H-2}$ and understanding the impact of the interference from BSs in the outer circle is one of the core contributions of this paper. For simplicity of representation, we let $I_r = \sum\nolimits_{i\in \Delta/b_0} g_i \hat{R}_i^{-\delta} + \sum\nolimits_{i\in \Delta_S} g_i \hat{R}_i^{-\delta} $ to denote the cumulative interference from all the other BSs $i$ (except the serving BS $b_0$) to the UE with the distance $\hat{R}_i$ and pathloss $g_i$. We could obtain the following lemma,
\begin{lemma}
\label{lemma:laplace_interference}
The Laplace transform of the cumulative interference $I_r$ for a cellular network with $\alpha$-stable distributed BSs could be formulated as
\begin{figure*}[!t]
\begin{equation}
\mathcal{L}_{I_r}(s) =
\left\{ \begin{aligned}
& \exp\Bigg\{  -\lambda \pi \mathbb{E}_g \Big( \Lambda(s,r,R) - \Theta(s,r,R)\Big)   -\frac{\sigma ^{\alpha} a^{\alpha (H-2)}}{\cos \frac{\pi \alpha}{2}} \Big[ \pi \mathbb{E}_g \Big(\Lambda(s,R,aR)\\
&\qquad - \Theta(s,R,aR)\Big)\Big]^{\alpha} -a^{H-2} \mu  \pi \mathbb{E}_g \Big(\Lambda(s,R,aR)- \Theta(s,R,aR)\Big) \Bigg\},  \alpha\neq 1;\\
&\exp\Bigg\{ \Big[2\sigma a^{H-2} \left( \ln \left\{ \pi \mathbb{E}_g \Big(\Lambda(s,R,aR)- \Theta(s,R,aR)\Big)\right\} +   (H-2) \ln a \right) - \pi a^{H-2}\mu\Big]  \\
& \qquad \cdot \mathbb{E}_g \Big(\Lambda(s,R,aR)- \Theta(s,R,aR)\Big)  -\lambda \pi \mathbb{E}_g \Big( \Lambda(s,r,R) - \Theta(s,r,R)\Big)  \Bigg\}, \alpha= 1.\\
\end{aligned} \right.
\end{equation}
\hrulefill
\vspace*{4pt}
\end{figure*}
where $\Theta(s,b,c) \stackrel{\textrm{def}}{=}  (s g)^{2/\delta} \left(\Gamma(-\frac{2}{\delta}+1, s g b^{-\delta}) - \Gamma(-\frac{2}{\delta} +1, s g c^{-\delta}) \right)  $. $\Gamma(d,x)=\int_{x}^{\infty} t^{d-1} e^{-t} \dif t$ denotes the incomplete Gamma function, while $\Gamma(d)=\int_{0}^{\infty} t^{d-1} e^{-t} \dif t$ denotes the standard Gamma function. Besides, $\Lambda(s,b,c)  \stackrel{\textrm{def}}{=}  c^2 \left[ 1 - \exp \left( -s g c^{-\delta} \right) \right] - b^2 \left[ 1 - \exp \left( -s g b^{-\delta} \right) \right]  $. 
\end{lemma}

We leave the proof of Lemma \ref{lemma:laplace_interference} in Appendix \ref{sec:proof_lemma_lapalace_interference} and further give the following theorem.
\begin{theorem}
\label{thm:coverage_prob}
The coverage probability in a cellular networks with $\alpha$-stable distributed BS density and the self-similarity is given by\newline
1) if $\alpha \neq 1$
\begin{align}
&p_c(T,\mathbb{S}(\alpha,\sigma,\mu),\delta)\nonumber\\
= & \int_{r>0} 2\pi r \exp \Big\{ -\frac{\sigma ^{\alpha} a^{\alpha (H-2)}}{\cos \frac{\pi \alpha}{2}} \left[ \Xi(\zeta T r^{\delta},R,aR)  \right]^{\alpha}  \\ 
& \quad -a^{H-2} \mu  \Xi(\zeta T r^{\delta},R,aR) - \mu\Xi( \zeta T r^{\delta},r,R) \nonumber\\
& \quad -\frac{\sigma^{\alpha} }{\cos\frac{\pi \alpha}{2} }  (\Xi( \zeta T r^{\delta},r,R) + \pi r^2)^{\alpha} -\mu \pi r^2 -\zeta T r^{\delta} N_0  \Big\}\nonumber\\
&\quad  \cdot \Big[\frac{\sigma^{\alpha} \alpha}{\cos\frac{\pi \alpha}{2} } (\Xi( \zeta T r^{\delta},r,R) + \pi r^2)^{\alpha-1} +\mu  \Big] \dif r\nonumber
\end{align}
2) if $\alpha = 1$
\begin{align}
&p_c(T,\mathbb{S}(\alpha,\sigma,\mu),\delta)\nonumber\\
= &  \int_{r>0} 2\pi r \exp \Bigg\{ \Big[ \frac{2\sigma a^{H-2}}{\pi} \big( \ln \left\{ \Xi(\zeta T r^{\delta},R,aR)  \right\} \nonumber\\
& \quad +   (H-2) \ln a \big) - a^{H-2}\mu\Big] \Xi(\zeta T r^{\delta},R,aR) \\
&\quad +  \frac{2\sigma \Xi( \zeta T r^{\delta},r,R) + 2\sigma \pi r^2 }{\pi}\ln(\Xi( \zeta T r^{\delta},r,R) + \pi r^2)  \nonumber \\
&\quad -\mu \Xi( \zeta T r^{\delta},r,R) -\mu\pi r^2 -\zeta T r^{\delta} N_0  \Bigg\} \nonumber \\
&\quad  \cdot  \Big[-\frac{2\sigma}{\pi} \Big(\ln(\Xi( \zeta T r^{\delta},r,R) + \pi r^2)+1\Big) + \mu\Big] \dif r \nonumber
\end{align}
where $\Xi(s,b,c)=\pi \mathbb{E}_g \left( \Lambda(s,b,c) - \Theta(s,b,c)\right) $.
\end{theorem}

\begin{proof}
From Lemma \ref{lemma:coverage_prob}, the coverage probability could be calculated as
\begin{equation}
\label{eq:coverage_prob_theorem}
\begin{aligned}
& p_c(T,\mathbb{S}(\alpha,\sigma,\mu),\delta) \\
=&\iint_{r>0,\lambda>0} e^{-\zeta T r^{\delta} N_0 }\mathcal{L}_{I_r}( \zeta T r^{\delta}) \mathbb{P}(r|\lambda) f(\lambda) \dif r\dif \lambda\\
\stackrel{(a)}{=}&\int_{r>0}  2\pi r e^{-\zeta T r^{\delta} N_0 } \mathcal{L}_{\Delta_s}( \zeta T r^{\delta}) \Upsilon(r)\dif r
\end{aligned}
\end{equation}
where the equation (a) is the direct result of Corollary \ref{cor:pdf_r} and Lemma \ref{lemma:laplace_interference} and $\Upsilon(r)$ could be formulated as \eqref{eq:upsilon_r} on Page \pageref{eq:upsilon_r}.
\begin{figure*}
\begin{align}
\label{eq:upsilon_r}
\Upsilon(r) = \int_{\lambda>0} \exp \left\{ -\lambda \pi \left[ \mathbb{E}_g \Bigg( \Lambda(\zeta T r^{\delta},r,R) - \Theta(\zeta T r^{\delta},r,R)\Bigg) + r^2 \right] \right\} \lambda  f(\lambda) \dif \lambda 
\end{align}	
\hrulefill
\vspace*{4pt}
\end{figure*}

By merging \eqref{eq:coverage_prob_theorem}, \eqref{eq:upsilon_r}, and Lemma \ref{lemma:laplace_interference} and applying the relationship of Laplace transformation $\int_{x} e^{-sx}x f(x)\dif x=-\frac{\dif \Phi(s)}{\dif s}$, we have the result.
\end{proof}

When we consider the whole region (i.e., the outer circle spanning from $R$ to $\infty$, or $a \rightarrow \infty$), we get the following corollary.
\begin{corollary}
\label{cor:a_infty}
When $a \rightarrow \infty$,  the coverage probability could be reduced to\\
1) if $\alpha \neq 1$
\begin{align}
&p_c(T,\mathbb{S}(\alpha,\sigma,\mu),\delta)\nonumber\\
= & \int_{r>0} 2\pi r \exp \Big\{ - \mu\Xi( \zeta T r^{\delta},r,R)  -\frac{\sigma^{\alpha} }{\cos\frac{\pi \alpha}{2} }  (\Xi( \zeta T r^{\delta},r,R)\nonumber\\
& \quad + \pi r^2)^{\alpha} -\mu \pi r^2 -\zeta T r^{\delta} N_0  \Big\}\\
&\quad   \cdot \Big[\frac{\sigma^{\alpha} \alpha}{\cos\frac{\pi \alpha}{2} } (\Xi( \zeta T r^{\delta},r,R) + \pi r^2)^{\alpha-1} + \mu  \Big] \dif r \nonumber
\end{align}
2) if $\alpha = 1$
\begin{align}
&p_c(T,\mathbb{S}(\alpha,\sigma,\mu),\delta) \nonumber \\
= &  \int_{r>0} 2\pi r \exp \Bigg\{ -\mu \Xi( \zeta T r^{\delta},r,R)  -\mu\pi r^2  -\zeta T r^{\delta} N_0  \nonumber \\
&\quad + \frac{2\sigma \Xi( \zeta T r^{\delta},r,R) + 2\sigma \pi r^2 }{\pi}\ln(\Xi( \zeta T r^{\delta},r,R) + \pi r^2)  \Bigg\}  \nonumber\\
&\quad  \cdot  \Big[\mu-\frac{2\sigma}{\pi} \Big(\ln(\Xi( \zeta T r^{\delta},r,R) + \pi r^2)+1\Big) \Big] \dif r
\end{align}
\end{corollary}
\begin{proof}
Firstly, we have 
\begin{align}
& \lim\limits_{a\rightarrow \infty} a^2 R^2 \left[ 1 - \exp \left( -\zeta T g (\frac{r}{aR})^{\delta} \right) \right] \nonumber\\
= & \lim\limits_{k\rightarrow 0^{+}} \frac{ R^2 \left[ 1 - \exp \left( -\zeta T g (\frac{r}{R})^{\delta} k^{\delta} \right) \right]}{k^2} \\
\stackrel{(a)}{=} & \lim\limits_{k\rightarrow 0^{+}} \frac{  -\zeta T g (\delta -1) \frac{r^{\delta}}{R^{\delta-2}} }{2} \exp \left( -\zeta T g (\frac{r}{R})^{\delta} k^{\delta} \right) k^{\delta-2} \nonumber \\
= & 0\nonumber
\end{align}
where the equation (a) comes from the l'H\^opital's Rule \cite{taylor_hospitals_1952} and applies the fact that the pathloss exponent $\delta \geq 2$.
	
Hence, as $a\rightarrow \infty$, we have $a^{H-2} \rightarrow 0$ for $H\in (0,1)$. So, we have \eqref{eq:a_infty_equation} on Page \pageref{eq:a_infty_equation}.
\begin{figure*}
\begin{equation}
\label{eq:a_infty_equation}
\begin{aligned}
a^{H-2}\Xi(\zeta T r^{\delta},R,aR)
&= a^{H-2} \pi \mathbb{E}_g \Bigg( - R^2 \left[ 1 - \exp \left( -\zeta T g (\frac{r}{R})^{\delta} \right) \right] -  (\zeta T g )^{2/\delta} r^2 \Gamma(-\frac{2}{\delta}+1, \zeta T g (\frac{r}{R})^{\delta}) \Bigg)\\
&\quad + a^{H-2} \pi \mathbb{E}_g \Bigg( a^2 R^2 \left[ 1 - \exp \left( -\zeta T g (\frac{r}{aR})^{\delta} \right) \right] +  (\zeta T g )^{2/\delta} r^2 \Gamma(-\frac{2}{\delta}+1, \zeta T g (\frac{r}{aR})^{\delta}) \Bigg)\\
&\rightarrow 0
\end{aligned}
\end{equation}

\begin{align}
\label{eq:thm_r_infty}
p_c(T,\mathbb{S}(\alpha,\sigma,\mu),\delta)  =  & \int_{r>0} 2\pi r\Bigg[\frac{\sigma^{\alpha} \sigma}{\cos\frac{\pi \alpha}{2} } \bigg(\pi \mathbb{E}_g \Big(\frac{2(\zeta T g)^{2/\delta}r^2}{\delta} \big[\Gamma(-\frac{2}{\delta}, \zeta T g ) -  \Gamma(-\frac{2}{\delta})\big]    \Big)\bigg)^{\alpha-1} + \mu  \Bigg] \cdot    \\
&\quad  \exp \Bigg\{  -\frac{\sigma^{\alpha} }{\cos\frac{\pi \alpha}{2} } \Bigg(\pi \mathbb{E}_g \Big(\frac{2(\zeta T g)^{2/\delta}r^2}{\delta} \big[\Gamma(-\frac{2}{\delta}, \zeta T g ) -  \Gamma(-\frac{2}{\delta})\big]    \Big)\Bigg)^{\alpha} -\zeta T r^{\delta} N_0 \nonumber\\
& \quad - \mu\pi \mathbb{E}_g \Bigg(\frac{2(\zeta T g)^{2/\delta}r^2}{\delta} \big[\Gamma(-\frac{2}{\delta}, \zeta T g ) -  \Gamma(-\frac{2}{\delta})\big]    \Bigg)   \Bigg\}  \dif r \nonumber
\end{align}

\begin{equation}
	\label{eq:thm_r_infty_1}
	\begin{aligned}
	p_c(T,\mathbb{S}(\alpha,\sigma,\mu),\delta) = & \int_{r>0} 2\pi r \Big[\mu - \frac{2\sigma}{\pi} \Bigg(\ln \Big( \pi  \mathbb{E}_g \Big(\frac{2(\zeta T g)^{2/\delta}r^2}{\delta} \big[\Gamma(-\frac{2}{\delta}, \zeta T g ) -  \Gamma(-\frac{2}{\delta})\big]    \Big)  \Big)+1\Bigg) \Big] \cdot   \\
	&\quad \exp \Bigg\{ -\zeta T r^{\delta} N_0  2\sigma \mathbb{E}_g \Big(\frac{2(\zeta T g)^{2/\delta}r^2}{\delta} \big[\Gamma(-\frac{2}{\delta}, \zeta T g ) -  \Gamma(-\frac{2}{\delta})\big]    \Big)   \ln\Bigg( \pi  \mathbb{E}_g \Big(\frac{2(\zeta T g)^{2/\delta}r^2}{\delta}\\
	&\quad \big[\Gamma(-\frac{2}{\delta}, \zeta T g ) -  \Gamma(-\frac{2}{\delta})\big]    \Big)  \Bigg)  
		-\mu \pi  \mathbb{E}_g \Big(\frac{2(\zeta T g)^{2/\delta}r^2}{\delta} \big[\Gamma(-\frac{2}{\delta}, \zeta T g ) -  \Gamma(-\frac{2}{\delta})\big]    \Big)    \Bigg\} \dif r
	\end{aligned}
	\end{equation}	
\hrulefill
\vspace*{4pt}
\end{figure*}

From Theorem \ref{thm:coverage_prob}, it can be observed that when $a\rightarrow \infty$, only the term 	$a^{H-2}\Xi(\zeta T r^{\delta},R,aR) =a^{H-2}\pi \mathbb{E}_g \Bigg( \Lambda(\zeta T r^{\delta},R,aR) - \Theta(\zeta T r^{\delta},R,aR)\Bigg)$ will be affected. We have the conclusion.
\end{proof}

When $R\rightarrow \infty$, we can further simplify the results to get more interesting insight.
\begin{theorem}
\label{thm:r_infty}
When $R \rightarrow \infty$, the coverage probability could be reduced to \eqref{eq:thm_r_infty} and \eqref{eq:thm_r_infty_1} (on Page \pageref{eq:thm_r_infty_1}) for $\alpha \neq 1$ and $\alpha = 1$, respectively.

\end{theorem}

We can have Theorem \ref{thm:r_infty} by following the lines to prove Corollary \ref{cor:a_infty} and applying the relationship $\Gamma(s,x)=(s-1)\Gamma(s-1,x)+x^{s-1}e^{-x}$ and $\Gamma(s)=(s-1)\Gamma(s-1)$.

Next, we state the coverage probability when extra constraints are imposed, that is, the spatial density in the inner circle is fixed (i.e., $\sigma=0$ and $\lambda =\mu$ in Theorem \ref{thm:r_infty}). We get the following corollary.
\begin{corollary}
The coverage probability in cellular networks with fixed spatial BS density is
\begin{align}
&p_c(T,\mathbb{S}(\alpha,0,\lambda),\delta) =  \int_{r>0} 2\pi \lambda r \exp \Bigg\{  -\zeta T r^{\delta} N_0 \\
&\quad - \lambda\pi \mathbb{E}_g \Bigg(\frac{2(\zeta T g)^{2/\delta}r^2}{\delta} \left[\Gamma(-\frac{2}{\delta}, \zeta T g ) -  \Gamma(-\frac{2}{\delta})\right]    \Bigg)   \Bigg\}  \dif r \nonumber
\end{align}
\end{corollary}
Hence, when the spatial density is fixed and equals $\lambda$, the self-similarity patterns no longer take effect and our result could be reduced to the well-recognized conclusions of HPPP obtained by J. G. Andrews \textit{et al.} in \cite{andrews_tractable_2011}.

Also, based on Theorem \ref{thm:r_infty}, we can obtain the following theorem.

\begin{theorem}
	\label{thm:r_infty_upper}
	For $\alpha \neq 1$, the coverage probability in \eqref{eq:thm_r_infty} has an upper bound, that is,
	\begin{align}
		& p_c(T,\mathbb{S}(\alpha,\sigma,\mu),\delta) \label{eq:r_infty_upper}\\
		\leq & \max_r \Big( 2\pi \mu    C(r) \Big) \cdot \int_{r>0}  A (T,\mathbb{S}(\alpha,\sigma,\mu),\delta) \dif r \nonumber
		% \stackrel{(a)}{\leq} &  \int_{r>0} 2\pi \mu    C(r) \dif r \sqrt{\int_{r>0}  A^2 (T,\mathbb{S}(\alpha,\sigma,\mu),\delta) \dif r }\nonumber
	\end{align}
	where $A (T,\mathbb{S}(\alpha,\sigma,\mu),\delta) = \exp \Big\{  -\frac{\sigma^{\alpha} }{\cos\frac{\pi \alpha}{2} } B^{\alpha} \Big\} \cdot \Big[\frac{\sigma^{\alpha} \alpha}{\mu \cos\frac{\pi \alpha}{2} } \cdot B^{\alpha-1} + 1  \Big] $, $B = \pi \mathbb{E}_g \Big(\frac{2(\zeta T g)^{2/\delta}r^2}{\delta} \cdot \big[\Gamma(-\frac{2}{\delta}, \zeta T g ) -  \Gamma(-\frac{2}{\delta})\big]   \Big)$, and $C(r) =r \exp \big\{  -\zeta T r^{\delta} N_0  - \mu B \big\} $.
\end{theorem}

\begin{proof}
	The result could be derived by applying the H\"older Inequality \cite{rogers_extension_1987}. As Theorem \ref{thm:r_infty} states
	\begin{align}
		& p_c(T,\mathbb{S}(\alpha,\sigma,\mu),\delta) \nonumber\\
		= & \int_{r>0} 2\pi \mu   A (T,\mathbb{S}(\alpha,\sigma,\mu),\delta) C(r) \dif r \\
		\stackrel{(a)}{\leq} &  \max_r \Big( 2\pi \mu    C(r) \Big) \cdot \int_{r>0}  A (T,\mathbb{S}(\alpha,\sigma,\mu),\delta) \dif r \nonumber
	\end{align}
	where (a) follows from the H\"older Inequality. 
\end{proof}

We have the following corollary to show that the upper bound get tighter as the SINR threshold increases.
\begin{corollary}
	\label{cor:upperbound_high_sinr}
	As the SINR threshold $T \rightarrow \infty$,
	\begin{align}
		& p_c(T,\mathbb{S}(\alpha,\sigma,\mu),\delta) \nonumber \\
		\rightarrow & \max_r \Big( 2\pi \mu    C(r) \Big) \cdot \int_{r>0}  A (T,\mathbb{S}(\alpha,\sigma,\mu),\delta) \dif r
	\end{align}
\end{corollary}

\begin{proof}
	The equation in \eqref{eq:r_infty_upper} holds if and only if for almost all $r$,  $C(r) = \max_r \Big( C(r) \Big)$ or $\frac{\partial C(r)}{\partial r} = 0$ \cite{rogers_extension_1987}.
	As $T \rightarrow \infty$, 
	\begin{align}
		& \frac{\partial C(r)}{\partial r}\\
		= & \exp \big\{  -\zeta T r^{\delta} N_0  - \mu B \big\} \Big( 1 + r \big\{ - \zeta T r ^{\delta - 1} \delta N_0 \\
		&\quad - \mu \pi \mathbb{E}_g \big[\frac{4 (\zeta T g)^{\frac{2}{\delta}}r}{\delta} B \big]  \big\} \Big) \\
		\stackrel{(a)}{\rightarrow} & 0 \nonumber
	\end{align}
	where, the equation $(a)$ can be derived from the l'H\^opital's Rule. Then we have the corollary.
\end{proof}

The following lemma gives a lower bound of $B$.
\begin{lemma}
	\begin{equation}
		B(r) \geq \pi r^2
	\end{equation}
\end{lemma}
\begin{proof}
	From the definition of $B$, we have 
	\begin{align}
		B & =\pi  \mathbb{E}_g \Bigg( \frac{2(\zeta T g)^{2/\delta}r^2}{\delta} \big[\Gamma(-\frac{2}{\delta}, \zeta T g ) -  \Gamma(-\frac{2}{\delta})\big]\Bigg) \nonumber\\
		& \stackrel{(a)}{=} \pi  \mathbb{E}_g \Bigg( r^2 \exp(-\zeta T g) + (\zeta T g)^{2/\delta} r^2 \big[ \Gamma(-\frac{2}{\delta}+1) \nonumber \\
		& \quad -\Gamma(-\frac{2}{\delta}+1,\zeta T g)  \big] \Bigg)\\
		& = \pi  \mathbb{E}_g \Bigg( r^2 \exp(-\zeta T g) + (\zeta T g)^{2/\delta} r^2 \int_0^{\zeta T g} t^{-\frac{2}{\delta}} e^{-t} \textrm{d}t  \Bigg)\nonumber  \\
		& \stackrel{(b)}{>}  \pi  \mathbb{E}_g \Bigg( r^2 \exp(-\zeta T g) + (\zeta T g)^{2/\delta} r^2 (\zeta T g)^{-\frac{2}{\delta}} \int_0^{\zeta T g}  e^{-t} \textrm{d}t  \nonumber\Bigg)\\
		& = \pi r^2 \nonumber
	\end{align}	
	where (a) follows $\Gamma(s,x)=(s-1)\Gamma(s-1,x)+x^{s-1}e^{-x}$ and $\Gamma(s)=(s-1)\Gamma(s-1)$. (b) comes from the observation that if $f(x) > g(x)$ in $x \in [c,d]$, then $\int_c^d f(x) \dif x>\int_c^d g(x) \dif x$ where $c=0$, $d= \zeta T g$, $f(x) = x^{-\frac{2}{\delta}} e^{-x}$, and $g(x) = (\zeta T g)^{-\frac{2}{\delta}} e^{-x}$.
\end{proof}

Next, we can have further approximation of $D \stackrel{\textrm{def}}{=} \int_{r>0}  A (T,\mathbb{S}(\alpha,\sigma,\mu),\delta) \dif r$, that is
\begin{lemma}
	\begin{equation}
		D \approx \int_{B>0}  A (T,\mathbb{S}(\alpha,\sigma,\mu),\delta) \dif B
	\end{equation}
\end{lemma}
Then, we can obtain the following theorem.
\begin{theorem}
	\label{thm:sigma}
	For $\alpha \in (0,1)$, the upper bound of coverage probability in Theorem \ref{thm:r_infty_upper} decreases along with the increase of $\sigma$.
\end{theorem}
Theorem \ref{thm:sigma} could be achieved by deriving $\frac{\partial D}{\partial \sigma}$ and proving $\frac{\partial D}{\partial \sigma} < 0 $ for $\alpha \in (0,1)$. The proof details could be found in Appendix \ref{sec:proof_thm_sigma}. Also, we can have a corollary concerning the relationship between the upper bound of coverage probability and $\alpha$.

\begin{corollary}
	\label{cor:alpha}
	For $\alpha \in (0,1)$, when  $\sigma$ is sufficiently large, the upper bound of coverage probability decreases with the increase of $\alpha$.
\end{corollary}
\begin{proof}
	Similar to the proof of Theorem \ref{thm:sigma}, after deriving $\frac{\partial A (T,\mathbb{S}(\alpha,\sigma,\mu),\delta)}{\partial \alpha}$, we have \eqref{eq:partial_D_alpha_full} on Page \pageref{eq:partial_D_alpha_full}.

	\begin{figure*}
	\begin{align}
		\label{eq:partial_D_alpha_full}
		\frac{\partial D}{\partial \alpha} = & \int_{B>0} \frac{\partial A (T,\mathbb{S}(\alpha,\sigma,\mu),\delta)}{\partial \alpha}  \dif B \nonumber \\
		= & \int_{B>0} \exp \Big\{  -\frac{\sigma^{\alpha} }{\cos\frac{\pi \alpha}{2} } B^{\alpha} \Big\} \cdot  \frac{(\sigma B)^{\alpha}}{\cos^2\frac{\pi \alpha}{2}} \cdot \Bigg( \ln(\sigma B) \big( (\frac{\alpha}{\mu B} - 1 )\cos \frac{\pi \alpha}{2} -\frac{\alpha}{\mu B}  (\sigma B)^{\alpha} \big) \nonumber \\
		&  - \sin \frac{\pi \alpha }{2} \cdot \frac{\pi \alpha }{2}  -\frac{1}{\mu B} \bigg\{(\sigma B)^{\alpha}  \tan \frac{\pi \alpha}{2}\cdot \frac{\pi \alpha}{2} - \cos \frac{\pi \alpha}{2} -  \sin \frac{\pi \alpha }{2} \cdot \frac{\pi \alpha}{2}\bigg\} \Bigg) \dif B 
	\end{align}	
	\hrulefill
	\vspace*{4pt}
	\end{figure*}

	For $\alpha \in (0,1)$ and a fixed $B$, when $\sigma$ is sufficiently large (i.e., $\sigma \gg 1$), we have $(\frac{\alpha}{\mu B} - 1 )\cos \frac{\pi \alpha}{2} -\frac{\alpha}{\mu B}  (\sigma B)^{\alpha} < 0$ and $(\sigma B)^{\alpha}  \tan \frac{\pi \alpha}{2}\cdot \frac{\pi \alpha}{2} - \cos \frac{\pi \alpha}{2} -  \sin \frac{\pi \alpha }{2} \cdot \frac{\pi \alpha}{2} > 0$. In other words, the integral item in \eqref{eq:partial_D_alpha_full} is negative for every $B$.
		
	So, the coverage probability decreases with the increase of $\alpha$.	
\end{proof}

\section{Numerical Analyses}
\label{sec:performance}

In this part, we provide numerical evaluations through which we compare the coverage probability of the cellular network with $\alpha$-stable distributed BS density and the HPPP model. Our simulation parameters are configured according to Table \ref{tb:notations}. In particular, the default spatial BS density for the HPPP model is consistent with that in \cite{andrews_tractable_2011}. As \cite{samorodnitsky_stable_1994} states that when $\alpha \in (0,1)$, the mean value of $\mathbb{S}(\alpha,\sigma,\mu)$ equals $\mu$. Therefore, it is fair to compare the $\alpha$-stable model and the single-tier HPPP model.

\begin{figure}
	\centering
	\includegraphics[width=0.475\textwidth]{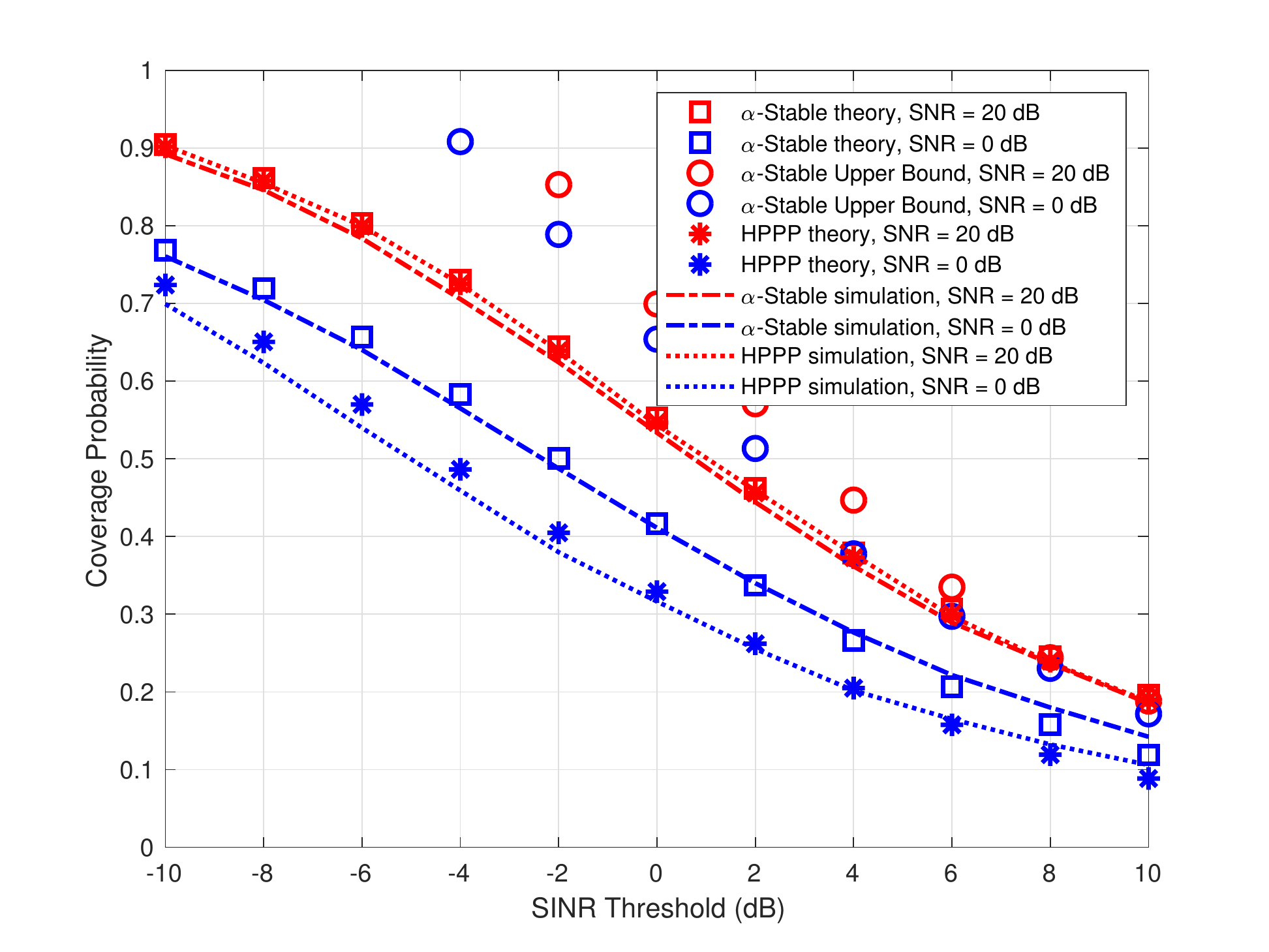}
	\vspace{-0.5cm}
	\caption{The coverage probability comparison under different SNR environment.}
	\label{fig:noise}
\end{figure}

\begin{figure*}
	\centering
	\includegraphics[width=0.875\textwidth]{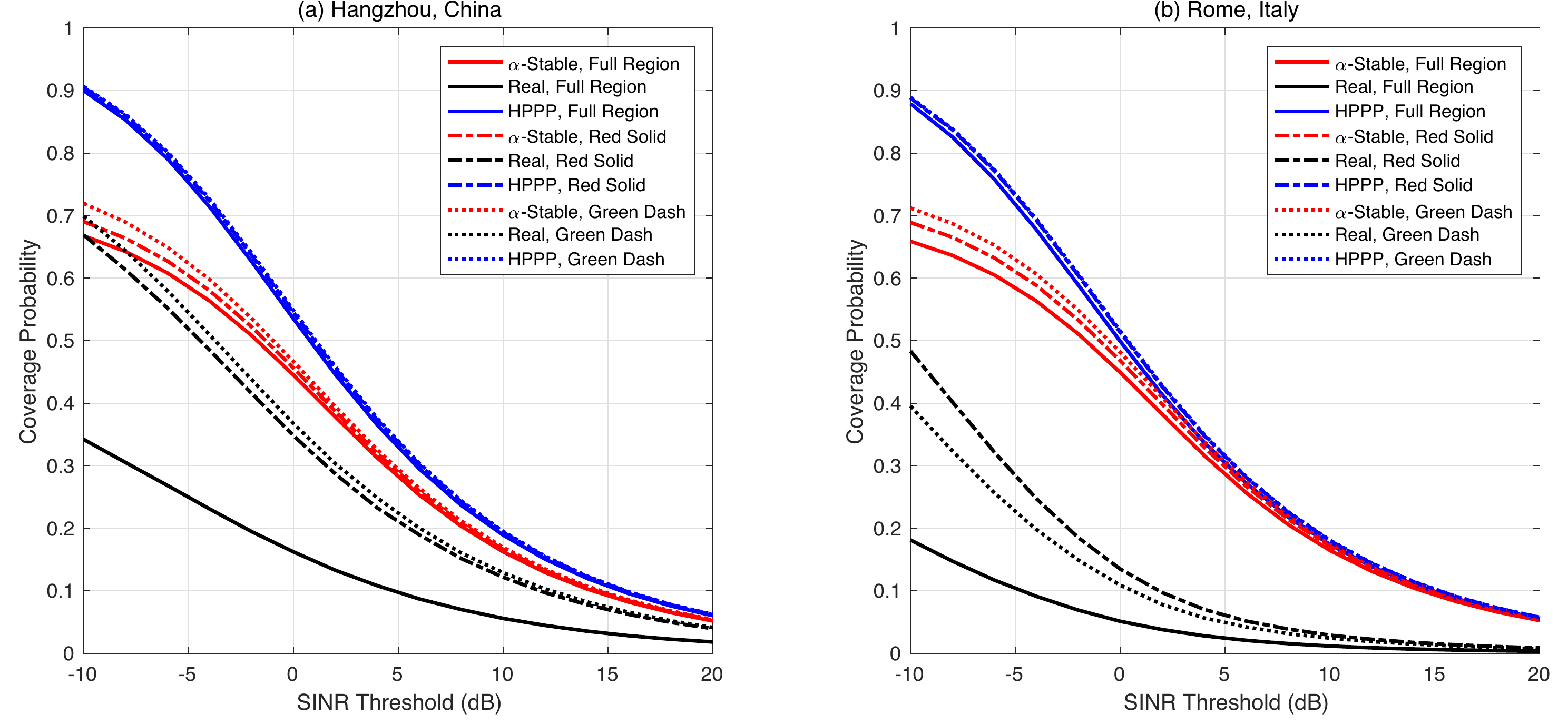}
	\vspace{-0.5cm}
	\caption{The coverage probability comparison between the models and the real environment for no noise cases, where ``Full Region" indicates the least square area to cover all BSs in Fig. \ref{fig:self-similarity-val}(a) (for Hangzhou) and Fig. \ref{fig:self-similarity-val}(d) (for Rome) in the revised manuscript while ``Red Solid" and ``Green Dash" refer to the selected square areas in these two subfigues.}
	\label{fig:real}
\end{figure*}

\begin{figure}
	\centering
	\includegraphics[width=0.475\textwidth]{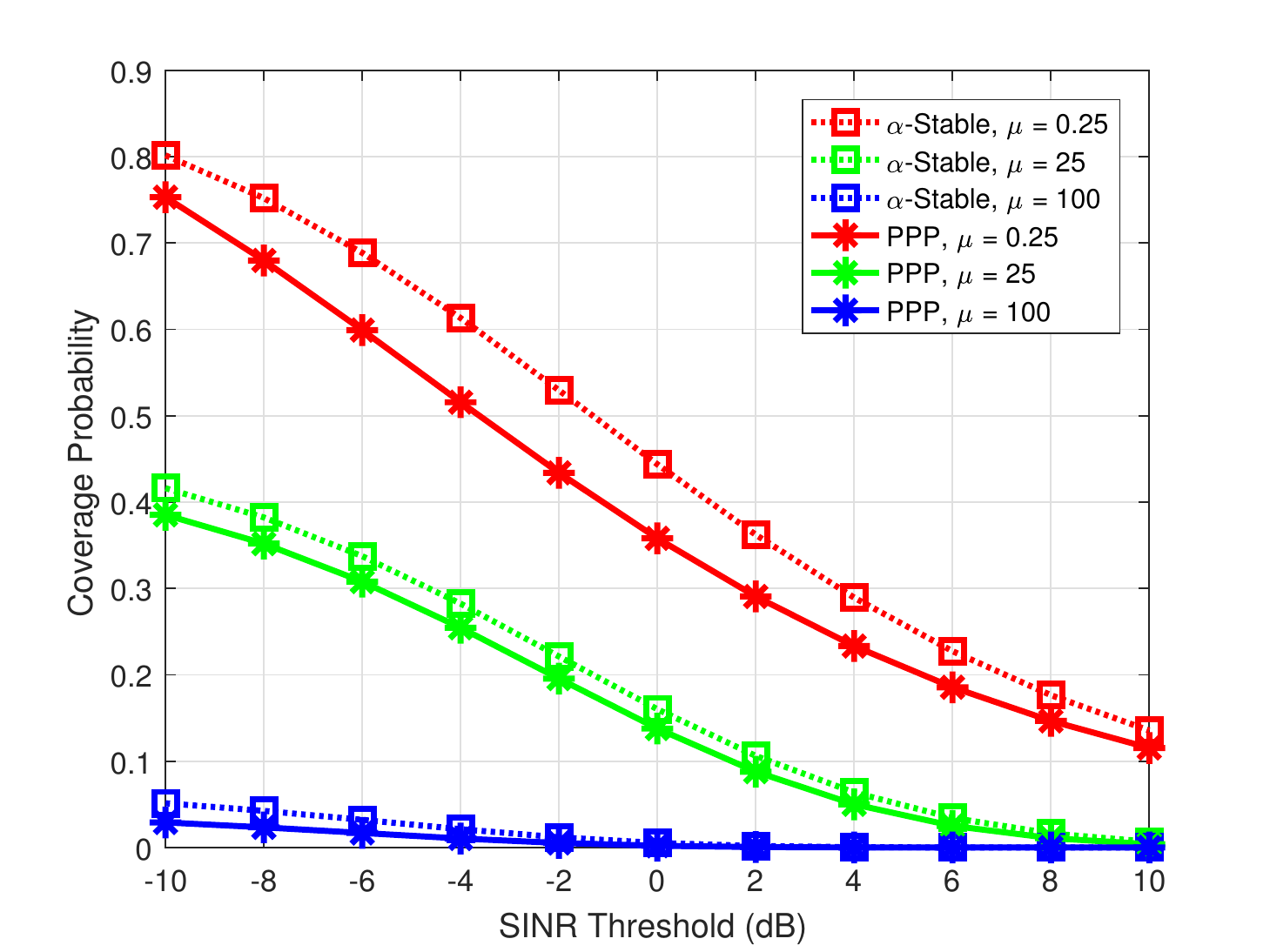}
	\vspace{-0.5cm}	
	\caption{The coverage probability comparison under various (average) spatial BS densities.}
	\label{fig:mu}
\end{figure}

\begin{figure}
	\centering
	\subfigure[The impact of $\sigma$\label{fig:sigma}]{\includegraphics[width=0.475\textwidth]{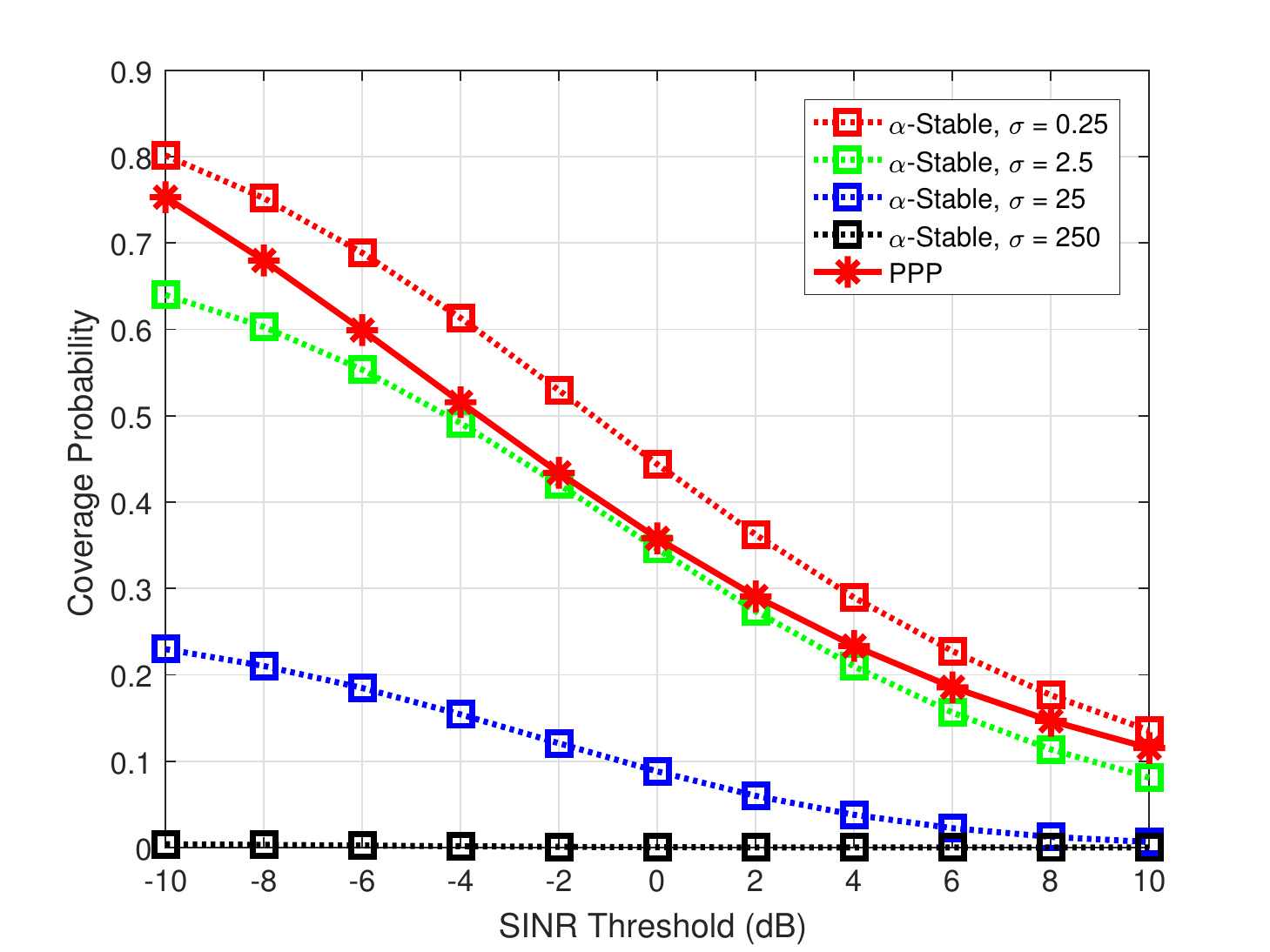}}	
	\subfigure[The impact of $\alpha$\label{fig:alpha}]{\includegraphics[width=0.475\textwidth]{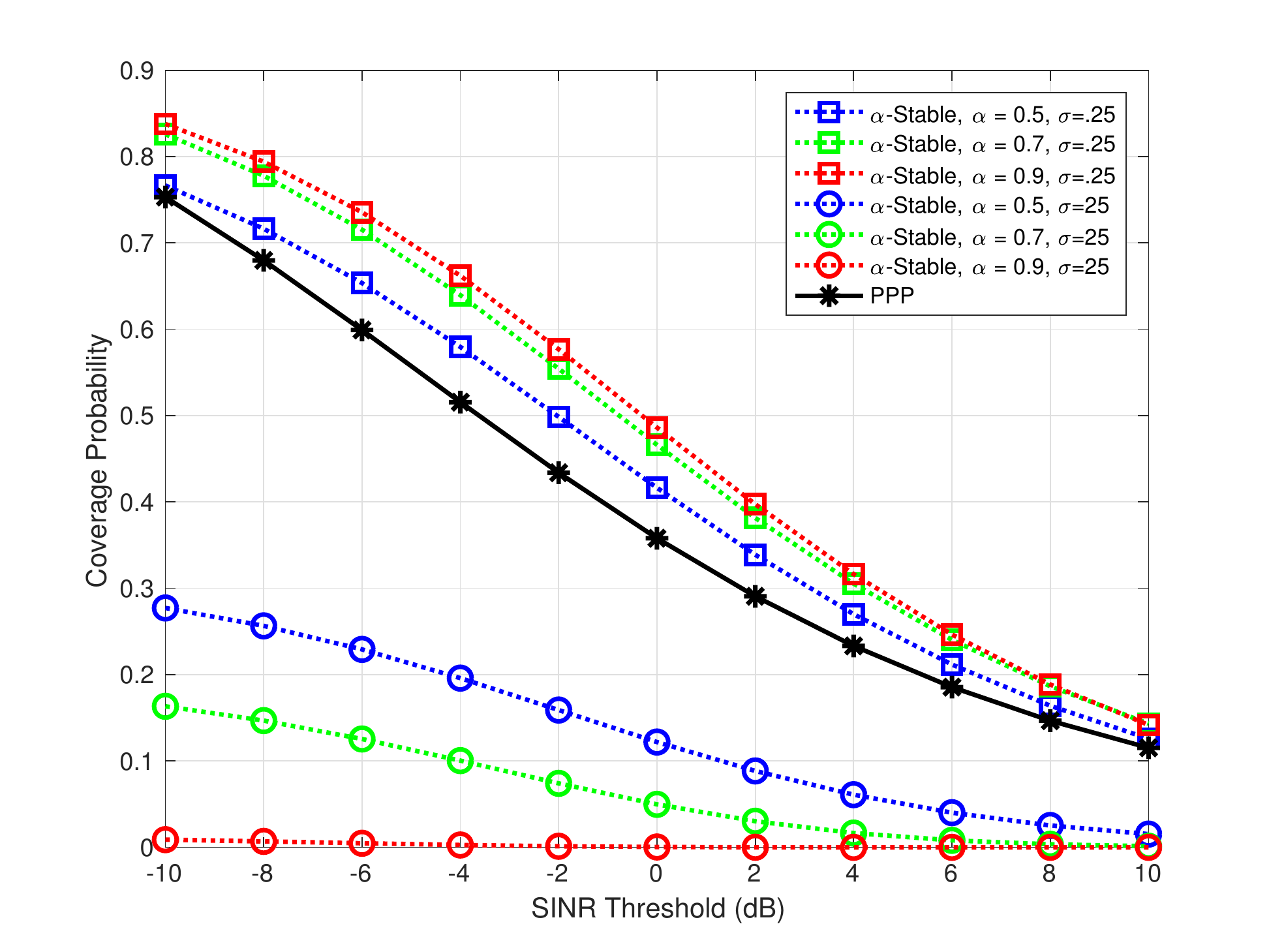}}
	\caption{Performance sensitivity analyses under different $\alpha$ and $\sigma$.}
\end{figure}

\begin{figure}
	\centering
	\subfigure[The impact of R\label{fig:r}]{\includegraphics[width=0.475\textwidth]{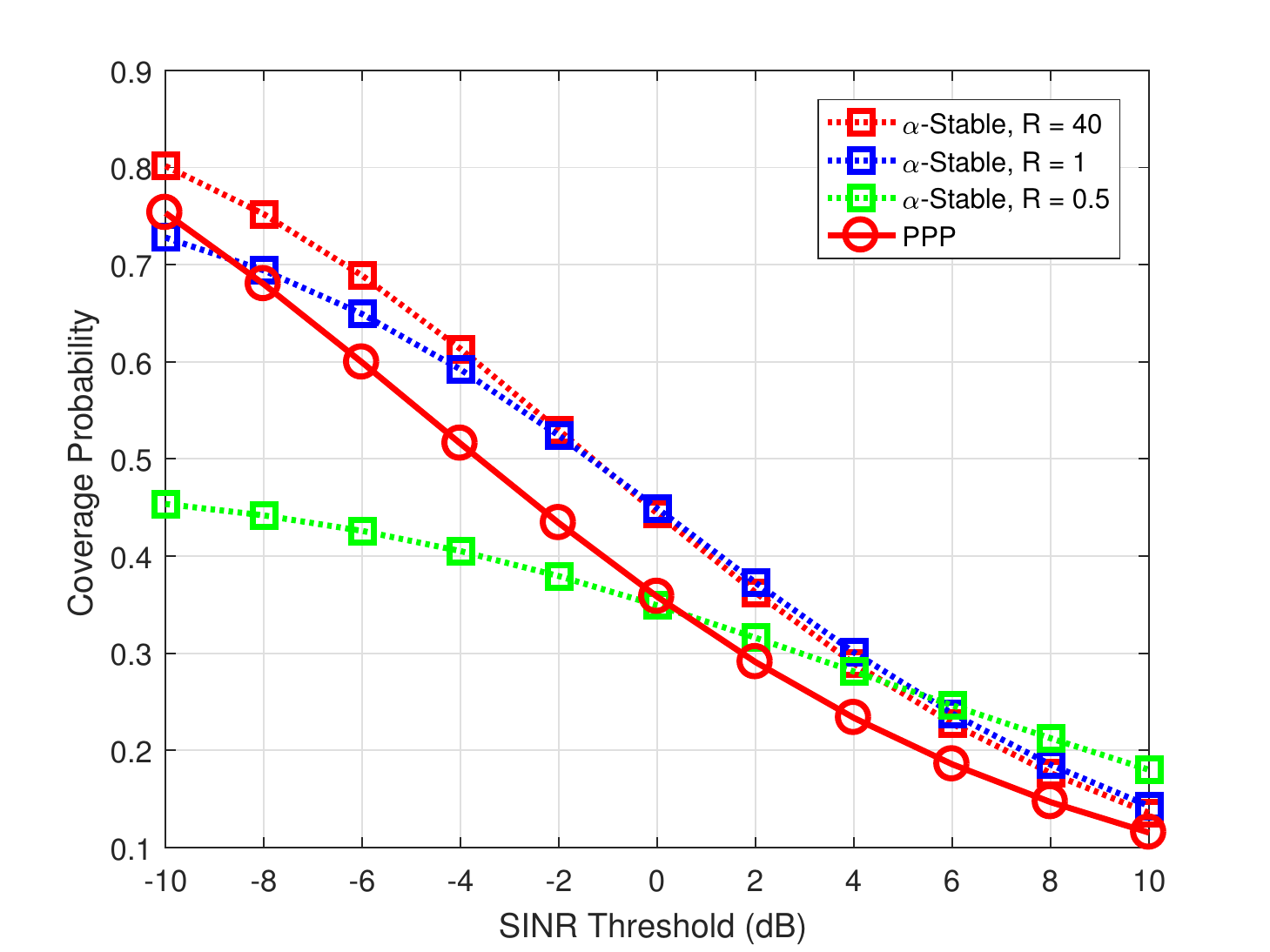}}
	\subfigure[The impact of $\delta$\label{fig:delta}]{\includegraphics[width=0.475\textwidth]{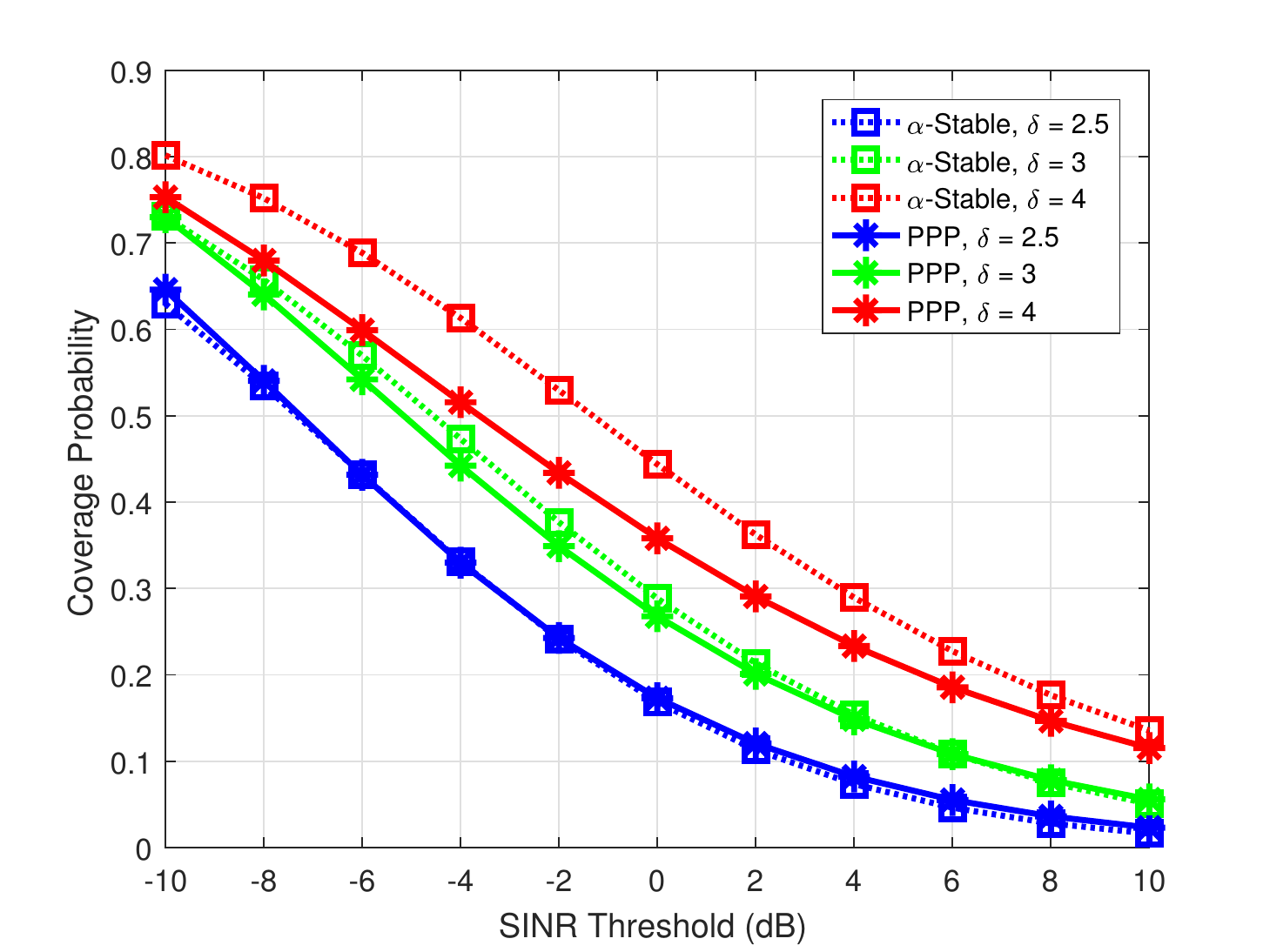}}
	\caption{Performance sensitivity analyses under different $R$ and $\delta$.}
\end{figure}

Firstly, we leverage the theoretical analysis results and give the coverage probability comparison under different AWGN environment in Fig. \ref{fig:noise}. From the figure, the performance gap between SNR $= 0$ dB and SNR $= 20$ dB is only $4$ dB and $6$ dB for the $\alpha$-stable model and the HPPP, respectively, which demonstrates that the cellular network is interference-limited rather than noise-limited. Also, there only exists acceptable performance gap between the theoretical analysis results and Monte Carlo simulations (see the dashed or dotted curves), where each curve is averaged under 15000 iterations. It further verifies the correctness of our theoretical derivations. Fig. \ref{fig:noise} also gives the upper bound of the coverage probability in Theorem \ref{thm:r_infty_upper} and shows the upper bound is quite close to the $\alpha$-stable model for high SINR thresholds, consistent with Corollary \ref{cor:upperbound_high_sinr}.

\begin{table*}
	\centering
	\caption{Fitting Parameters for Fig. \ref{fig:real}.}
	\begin{tabular}{ccccccc}
	\toprule
	\multirow {2}{*}{City} & \multirow {2}{*}{Area} & \multicolumn{4}{c}{$\alpha$-Stable} & Poisson \\	
	 &  & $\alpha$ & $\beta$ & $\sigma$ & $\mu$ & $\lambda$ \\
	\midrule
	\multirow {3}{*}{Hangzhou} & Full Region & 0.5298 & 1 & 6.955e-08 & 2.050e-07 & 1.117e-06\\
	& Red Solid & 0.5780 & 1 & 1.143e-07 & 1.527e-07 & 1.403e-06\\
	& Green dashed & 0.5641 & 1 & 1.822e-07 & 1.306e-07 & 1.890e-06 \\
	\midrule
	\multirow {3}{*}{Rome} & Full Region & 0.6830 & 1 & 6.422e-08 & 6.487e-07 & 6.487e-07\\
	& Red Solid & 0.7696 & 1 & 8.784e-08 & 9.290e-08 & 7.702e-07\\
	& Green dashed & 0.7442 & 1  & 8.438e-08 & 1.570e-07 & 7.928e-07 \\
	\bottomrule
	\end{tabular}%
	\label{tb:paraEst}%
	\end{table*}%

Fig. \ref{fig:real} gives the coverage probability comparison between the models and the real environment. For the models, we estimate the unknown parameters based on the methodology in \cite{zhou_alpha--stable_2015} and obtain the coverage probability based on these estimated parameters summarized in Table \ref{tb:paraEst}. For the real environment, we randomly drop the users in 100000 iterations and count the coverage probability after calculating the SINR for each random drop. Notably, the ``full region" indicates the least square surrounding all BSs in Hangzhou and Rome in Fig. \ref{fig:self-similarity-val}. Since the ``full region'' contains some areas of other administrative areas with unknown BS deployment information, the deduced coverage probability in Fig. \ref{fig:real} is rather low. Hence, we also choose two smaller regions (i.e., those areas within red solid line and green dash line). From the subfigures, compared with the single-tier HPPP model, the $\alpha$-stable model could significantly approximate the real environment, especially for lower SINR thresholds.
	
Fig. \ref{fig:mu} illustrates interesting coverage probability comparison under various (average) spatial BS densities. It shows that $\mu = 0.25$ has lead to remarkable coverage probability and also made the networks to be interference-limited. Therefore, further increasing the spatial density would add more to the aggregated interference and impose negative impact on the coverage probability. It also implies that the influence of the spatial BS density on the coverage probability is rather sophisticated. Fig. \ref{fig:sigma} and Fig. \ref{fig:alpha} further examine the impact of $\sigma$ and $\alpha$. Consistent with the theoretical finds in Theorem \ref{thm:sigma}, a larger $\sigma$ leads to worse coverage probability. On the other hand, Fig. \ref{fig:noise}, Fig. \ref{fig:real}, and Fig. \ref{fig:mu} seem to imply that for high SINR thresholds, the gap between the $\alpha$-stable model and the HPPP becomes smaller. But, Fig. \ref{fig:sigma} shows that such an observation only holds for small $\sigma$ values, since for high SINR thresholds, a sufficiently large received signal from the nearest serving BS plays the determinant role and the impact of all the other interfering BSs becomes comparatively smaller. In this regard, Corollary \ref{cor:pdf_r} shows that small $\sigma$ values produce similar results for the PDF of the distance from the nearest BS to the target UE. Hence, Fig. \ref{fig:sigma} depicts that small $\sigma$ values lead to small gap between $\alpha$-stable model and the HPPP but the performance differences still hold when $\sigma = 25$ or $\sigma = 250$.  From Fig. \ref{fig:alpha}, as implied by Corollary \ref{cor:alpha}, when $\sigma = 25$, a larger $\alpha$ incurs inferior coverage probability. However, the interesting phenomenon is that when $\sigma = 0.25$, a larger $\alpha$ incurs superior coverage probability. 

\begin{table*}
	\centering
	\caption{A summary of the network coverage probability under different $a$ and $H$.}
	\label{tb:a_h}
	\begin{tabular}{r|lllll|l}
		\toprule
		\multirow{3}{*}{SNR} & \multicolumn{5}{c|}{$\alpha$-Stable} & \multirow{3}{*}{HPPP}\\
		 & $a = 2$ & $a = 20$ & 
		$a = 200$ &
		$a = 2$ &
		$a = 2$ &
		\\
		& $H = 0.9$ & $H = 0.9$ & 
		$H = 0.9$ &
		$H = 0.1$ &
		$H = 0.5$ & \\
		\midrule                              
		-10 dB & 0.8015 & 0.8019 & 0.8020 & 0.8016 & 0.8016 & 0.7531 \\
		\hline                                              
		0 dB & 0.4437 & 0.4441 & 0.4443 & 0.4439 & 0.4438 & 0.3580 \\
		\hline                                              
		10 dB & 0.1355 & 0.1357 & 0.1357 & 0.1356 & 0.1355 & 0.1153 \\
		\bottomrule
	\end{tabular}
\end{table*}

Next, we focus on the impact of the spatial self-similarity on the coverage probability. Recalling the statements in Section \ref{sec:self-similarity}, the spatial self-similarity is examplified by the relationship $N(aR) \sim a^H N(R)$ or $\lambda_S \sim a^{H-2} \lambda$. Therefore, we examine the coverage probability when $a$ and $H$ differs and provide the corresponding results in Table \ref{tb:a_h}. From the table, we observe increasing $H$ or decreasing $a$ will result in a slight reduction of the coverage probability. The trivial impact of $a$ and $H$ on the network performance can be explained as that a variation of $H$ or $a$ will make the spatial BS density $\lambda_S$ of the outer circle larger from a probabilistic sense, thus making BSs over-crowded and generating huge interference. This phenomenon that stronger self-similarity incurs negative impact is also consistent with its counterpart of self-similar Ethernet traffic. As stated in \cite{park_effect_1997}, stronger self-similarity in Ethernet traffic will make heavier traffic prone to arrive in a sequel and congest the network.

We continue the performance analyses in Fig. \ref{fig:r} and Fig. \ref{fig:delta}. Fig. \ref{fig:r} shows reducing $R$ will cause a reduction of coverage probability for low SINR thresholds but an increment for high SINR thresholds. On the other hand, consistent with our intuition, Fig. \ref{fig:delta} depicts that a smaller pathloss exponent will bring a lower coverage probability.

\section{Conclusion}
\label{sec:conclusion}
In this paper, we have performed the stochastic geometry analyses in cellular networks with $\alpha$-stable distributed BS density. By leveraging the self-similarity among BSs, which is verified based on the practical BS deployment records in China and Europe (Italy), we have provided a tractable solution for the coverage probability in a large-scale area. We have demonstrated that our analytical results could be reduced to the works achieved by J. G. Andrews \cite{andrews_tractable_2011}. We have also given an upper bound for high SINR thresholds and theoretically shown the monotonicity of this bound with respect to the variance of BS density and validates that a larger variance of BS density leads to smaller coverage probability. Besides, we have simulated the coverage probability performance under extensive parameter settings and verified the consistence between the theoretical and simulation results. Our results have shown that compared to the single-tier HPPP model, the $\alpha$-stable model yields closer performance to the real environment especially for lower SINR thresholds and adds to the deeper understanding of the impact of BS densities on the coverage probability, by incorporating the stability parameter $\alpha$ and the scale factor $\sigma$ to characterize the BS deployment inhomogeniety. Therefore, it could contribute to analyze the performance under more sophisticated network configurations and make it easier to understand the actual network variations.

There still exist some open questions to be addressed. For example, due to the lack of two-tier or multiple-tier BS deployment information, we can not extract the relevant fitting parameters behind two-tier or multiple-tier HPPP models. Thus, our work only compare the $\alpha$-stable model with single-tier HPPP models. Actually, the comparison with two-tier or multiple-tier HPPP models is very interesting. Also, it is quite important to further study how to obtain an even more computational efficient approximation for the coverage probability of the $\alpha$-stable model.
Besides, our work has shown that instead of improving the coverage performance, simple yet stubborn deployment of BSs incurs significant interference and degrades the coverage performance. Therefore, frequency reuse has been applied in practical cellular communication. In this case, it is still meaningful to combine $\alpha$-stable self-similarity with more realistic network configurations, so as to produce more valuable results. 
\appendix

\subsection{The Proof of Lemma \ref{lemma:laplace_interference}}
\label{sec:proof_lemma_lapalace_interference}
\begin{proof}
According to the definition of Laplace transform, we have \eqref{eq:laplace_decompose_ir} on Page \pageref{eq:laplace_decompose_ir}, where the equation (a) in \eqref{eq:laplace_decompose_ir} comes from the identical, independent distribution of $g_i$ and its further independence from the point process.
\begin{figure*}
\begin{align}
\label{eq:laplace_decompose_ir}
&\mathcal{L}_{I_r}(s) =\mathbb{E}_{I_r} \left[e^{-sI_r} \right]=\mathbb{E}_{\Delta+\Delta_s,g_i} \left[\exp \left( -s\sum\limits_{i\in (\Delta+\Delta_s)/b_0} g_i \hat{R}_i^{-\delta} \right)\right]\\
&\stackrel{(a)}{=} \mathbb{E}_{\Delta} \left[\prod_{i\in \Delta/b_0} \mathbb{E}_g \left[ \exp \left( -s g \hat{R}_i^{-\delta} \right) \right] \right]\cdot \mathbb{E}_{\Delta_S} \left[\prod_{i\in \Delta_S} \mathbb{E}_g \left[ \exp \left( -s g \hat{R}_i^{-\delta} \right) \right] \right] \nonumber
\end{align}	
\hrulefill
\vspace*{4pt}
\end{figure*}

From \eqref{eq:laplace_decompose_ir}, it can be observed that the Laplace transform is composed of two parts (i.e., $\mathcal{L}_{\Delta}(s) =\mathbb{E}_{\Delta} \left[\prod_{i\in \Delta/b_0} \mathbb{E}_g \left[ \exp \left( -s g \hat{R}_i^{-\delta} \right) \right] \right] $ and $\mathcal{L}_{\Delta_S}(s) =\mathbb{E}_{\Delta_S} \left[\prod_{i\in \Delta_S} \mathbb{E}_g \left[ \exp \left( -s g \hat{R}_i^{-\delta} \right) \right] \right]  $), which is also consistent with our intuition. Next, we derive the representation of these two parts separately.

\begin{itemize}
\item For $\mathcal{L}_{\Delta}(s)$, we have a fixed spatial density $\lambda$. Then, from the probability generating functional (PGFL) for the PPP \cite{chiu_stochastic_2013}, we have 
\begin{displaymath}
    \mathbb{E}\left[ \prod_{x\in\Delta} f(x)\right] = \exp \left( -\lambda \int_{\Delta}(1-f(x)) \dif x\right).
\end{displaymath}

Since $f(x) = \mathbb{E}_g \left[ \exp \left( -s g \hat{R}^{-\delta} \right) \right]$, we have
\begin{align}
\label{eq:integral_1_f(x)_delta}
&\int_{\Delta}(1-f(x)) \dif x \nonumber\\
= & 2\pi \int_{r}^{R} \left\{ 1 -  \mathbb{E}_g \left[ \exp \left( -s g v^{-\delta} \right) \right] \right\} v \dif v \\
\stackrel{(a)}{=} & \pi \mathbb{E}_g \Big\{\Lambda(s,r,R)- \Theta(s,r,R)\Big\} \nonumber
\end{align}

We leave the derivation of the equation (a) in Appendix \ref{app:integral_1_f(x)_delta}. Accordingly, we have 
\begin{equation}
\label{eq:laplace_delta}
    \mathcal{L}_{\Delta}(s) =  \exp \left[ -\lambda \pi \mathbb{E}_g \Bigg( \Lambda(s,r,R) - \Theta(s,r,R)\Bigg)  \right].
\end{equation}
\item Similarly, for $\mathcal{L}_{\Delta_S}(s)$, as $\lambda_S \sim \lambda a^{H-2}$, we have the equation for $\mathcal{L}_{\Delta_S}(s)$ by using the Laplace transform of $\lambda_S$ in Lemma  \ref{lemma:laplace_lambda_s} and applying the result $\int_{\Delta_S}(1-f(x)) \dif x = \pi \mathbb{E}_g \big\{\Lambda(s,R,aR)- \Theta(s,R,aR)\big\}$, which can be obtained by adopting a similar methodology as the derivation of \eqref{eq:integral_1_f(x)_delta}. Specifically, \eqref{eq:laplace_delta_s} on Page \pageref{eq:laplace_delta_s} can be obtained, where the equation (a) therein is achieved by using the Laplace transform of $\lambda_S$ in Lemma  \ref{lemma:laplace_lambda_s} and applying the result $\int_{\Delta_S}(1-f(x)) \dif x = \pi \mathbb{E}_g \left\{\Lambda(s,R,aR)- \Theta(s,R,aR)\right\}$, which can be obtained by adopting a similar methodology as the derivation of \eqref{eq:integral_1_f(x)_delta}.
\end{itemize}

\begin{figure*}
\begin{align}
\label{eq:laplace_delta_s}
&\mathcal{L}_{\Delta_S}(s) =\int_{\lambda_S=0}^{\infty} \exp \left( -\lambda_S \int_{\Delta_S}(1-f(x)) \dif x\right)    f(\lambda_S) \dif \lambda_S \nonumber\\
&\stackrel{(a)}{=}\left\{ \begin{aligned}
& \exp\Bigg\{ -\frac{\sigma ^{\alpha} a^{\alpha (H-2)}}{\cos \frac{\pi \alpha}{2}} \left[ \pi \mathbb{E}_g \Bigg(\Lambda(s,R,aR)- \Theta(s,R,aR)\Bigg)\right]^{\alpha}\\
&\qquad  -a^{H-2} \mu  \pi \mathbb{E}_g \Bigg(\Lambda(s,R,aR)- \Theta(s,R,aR)\Bigg)\Bigg\},  \alpha\neq 1;\\
&\exp\Bigg\{ \Bigg[2\sigma a^{H-2} \left( \ln \left\{ \pi \mathbb{E}_g \Bigg(\Lambda(s,R,aR)- \Theta(s,R,aR)\Bigg)\right\} +   (H-2) \ln a \right) \\
& \qquad - \pi a^{H-2}\mu\Bigg] \cdot \mathbb{E}_g \Bigg(\Lambda(s,R,aR)- \Theta(s,R,aR)\Bigg) \Bigg\}, \alpha= 1.\\
\end{aligned} \right.
\end{align}	
\hrulefill
\vspace*{4pt}
\end{figure*}

Combining \eqref{eq:laplace_delta} and \eqref{eq:laplace_delta_s}, we obtain the result.
\end{proof}

\subsection{Derivations of \eqref{eq:integral_1_f(x)_delta}}
\label{app:integral_1_f(x)_delta}
In this part, we provide the details about the equation (a) in the \eqref{eq:integral_1_f(x)_delta}.
\begin{align}
\label{eq:derivation_laplace_1_delta}
&\int_{\Delta}(1-f(x)) \dif x \nonumber  \\
& =  2\pi \int_{r}^{R} \left\{ 1 -  \mathbb{E}_g \left[ \exp \left( -s g v^{-\delta} \right) \right] \right\} v \dif v \nonumber  \\
& = 2\pi \int_{0}^{\infty} \int_{r}^{R} \left\{ 1 - \exp \left( -s g v^{-\delta} \right) \right\} v \dif v	f(g) \dif g \\
& = \pi \int_{0}^{\infty} \underbrace{\int_{r^{-\delta}}^{R^{-\delta}} \left\{ 1 - \exp \left( -s g y \right) \right\}  \dif y^{-\frac{2}{\delta}} }_{(\divideontimes)}	f(g) \dif g \nonumber  \\
&= \pi \mathbb{E}_g \Big\{ \Lambda(s,r,R)- \Theta(s,r,R)\Big\} \nonumber
\end{align}
The last equation holds, since for the inside integral $(\divideontimes)$, we have
\begin{align}
& (\divideontimes) \nonumber \\
= & - (sg)^{\frac{2}{\delta}} \int_{r^{-\delta}}^{R^{-\delta}}  (sgy)^{-\frac{2}{\delta}}\exp \left( -s g y \right) \dif y \nonumber\\
&\quad + y^{-\frac{2}{\delta}}\left[ 1 - \exp \left( -s g y \right) \right] \Big|_{r^{-\delta}}^{R^{-\delta}}  \\
= & R^2 \left[ 1 - \exp \left( -s g R^{-\delta} \right) \right] - r^2 \left[ 1 - \exp \left( -s g r^{-\delta} \right) \right] \nonumber \\
& \quad - (sg)^{\frac{2}{\delta}} \left[\Gamma(-\frac{2}{\delta} +1, sg r^{-\delta}) -  \Gamma(-\frac{2}{\delta} +1, sg R^{-\delta}) \right] \nonumber
\end{align}

\subsection{The Proof of Theorem \ref{thm:sigma}}
\label{sec:proof_thm_sigma}
\begin{proof}
	
By deriving $\frac{\partial A (T,\mathbb{S}(\alpha,\sigma,\mu),\delta)}{\partial \sigma}$, we have
\begin{align}
	& \frac{\partial A (T,\mathbb{S}(\alpha,\sigma,\mu),\delta)}{\partial \sigma}  \nonumber \\
	= & \exp \Big\{  -\frac{\sigma^{\alpha} }{\cos\frac{\pi \alpha}{2} } B^{\alpha} \Big\} \cdot \Big[\frac{\sigma^{\alpha} \alpha }{\mu \cos\frac{\pi \alpha}{2} } \cdot B^{\alpha-1} + 1  \Big] \cdot \Big[-\frac{\sigma^{\alpha-1}\alpha}{\cos\frac{\pi \alpha}{2} } B^{\alpha} \Big] \nonumber \\
	& \quad +  \exp \Big\{  -\frac{\sigma^{\alpha} }{\cos\frac{\pi \alpha}{2} } B^{\alpha} \Big\} \cdot \Big[\frac{\alpha^2 \sigma^{\alpha -1}}{\mu \cos\frac{\pi \alpha}{2} } \cdot B^{\alpha-1}  \Big] \label{eq:partial_sigma}\\
	= & \exp \Big\{  -\frac{\sigma^{\alpha} }{\cos\frac{\pi \alpha}{2} } B^{\alpha} \Big\} \cdot \Big[\frac{\alpha \sigma^{\alpha -1}}{\mu \cos\frac{\pi \alpha}{2} } \cdot B^{\alpha-1}  \Big] \nonumber\\
	&\quad \cdot \Big[\alpha - \mu  B - \frac{\sigma^{\alpha} \alpha}{\cos\frac{\pi \alpha}{2} } \cdot B^{\alpha}  \Big] \nonumber
\end{align}

\begin{figure*}
\begin{align}
	\label{eq:proof_thm_sigma_2}
	& \frac{\partial D}{\partial \sigma} = \int_{B>0} \frac{\partial A (T,\mathbb{S}(\alpha,\sigma,\mu),\delta)}{\partial \sigma}  \dif B \nonumber \\
	= & \int_{B>0} \exp \Big\{  -\frac{\sigma^{\alpha}}{\cos\frac{\pi \alpha}{2} } B^{\alpha} \Big\} \cdot \Big[\frac{\alpha \sigma^{\alpha}}{ \cos\frac{\pi \alpha}{2} } \cdot B^{\alpha-1}  \Big] \cdot \Big[\frac{\alpha}{\mu\sigma} - \frac{B}{\sigma} - \frac{\sigma^{\alpha-1} \alpha }{\mu \cos\frac{\pi \alpha}{2} } \cdot B^{\alpha}  \Big] \dif B \nonumber \\
	= &  - \frac{\alpha}{\mu\sigma} \exp \Big\{  -\frac{\sigma^{\alpha} }{\cos\frac{\pi \alpha}{2} } B^{\alpha} \Big\} \Bigg|_0^{\infty} +\frac{B}{\sigma} \exp \Big\{  -\frac{\sigma^{\alpha}}{\cos\frac{\pi \alpha}{2} } B^{\alpha} \Big\} \Bigg|_0^{\infty}  \nonumber   \\
	& - \frac{1}{\sigma} \int_{B>0} \exp \Big\{  -\frac{\sigma^{\alpha}}{\cos\frac{\pi \alpha}{2} } B^{\alpha} \Big\}   \dif B + \frac{\sigma^{\alpha-1} \alpha}{\mu \cos\frac{\pi \alpha}{2} } \cdot B^{\alpha} \exp \Big\{  -\frac{\sigma^{\alpha}}{\cos\frac{\pi \alpha}{2} } B^{\alpha} \Big\} \Bigg|_0^{\infty} \\
	& 
	- \int_{B>0} \exp \Big\{  -\frac{\sigma^{\alpha}}{\cos\frac{\pi \alpha}{2} } B^{\alpha} \Big\} \dif \Big(\frac{\sigma^{\alpha-1} \alpha }{\mu \cos\frac{\pi \alpha}{2} } \cdot B^{\alpha}\Big)\nonumber\\
	\stackrel{(a)}{=} & \frac{\alpha}{\mu \sigma } - \frac{1}{\sigma} \int_{B>0} \exp \Big\{  -\frac{\sigma^{\alpha}}{\cos\frac{\pi \alpha}{2} } B^{\alpha} \Big\}   \dif B + \frac{\alpha}{\mu\sigma} \exp \Big\{  -\frac{\sigma^{\alpha} \alpha}{\cos\frac{\pi \alpha}{2} } B^{\alpha} \Big\} \Bigg|_0^{\infty}\nonumber \\
	< & 0 \nonumber
\end{align}
\hrulefill
\vspace*{4pt}
\end{figure*}

For $\alpha \in (0,1)$, \eqref{eq:proof_thm_sigma_2} on Page \pageref{eq:proof_thm_sigma_2} can be obtained, where (a) in \eqref{eq:proof_thm_sigma_2} comes from the fact that 
$\exp \Big\{  -\frac{\sigma^{\alpha}}{\cos\frac{\pi \alpha}{2} } B^{\alpha} \Big\} \Bigg|_0^{\infty}  = -1$. Due to the l'H\^opital's Rule, $\lim_{B \rightarrow \infty} B^{\alpha} \exp \Big\{  -\frac{\sigma^{\alpha}}{\cos\frac{\pi \alpha}{2} } B^{\alpha} \Big\} = \lim_{B \rightarrow \infty} \frac{\alpha B^{\alpha -1}}{\exp \Big\{ \frac{\sigma^{\alpha}}{\cos\frac{\pi \alpha}{2} } B^{\alpha} \Big\}  \frac{\sigma^{\alpha} \alpha}{\cos\frac{\pi \alpha}{2} } B^{\alpha-1} }  =  0 $. Also, by iteratively applying the l'H\^opital's Rule, we can have
\begin{align}
	& \lim_{B \rightarrow \infty} B \exp \Big\{  -\frac{\sigma^{\alpha}}{\cos\frac{\pi \alpha}{2} } B^{\alpha} \Big\} \nonumber\\
	= & \lim_{B \rightarrow \infty} \frac{B^{1-\alpha} }{\exp \Big\{ \frac{\sigma^{\alpha}}{\cos\frac{\pi \alpha}{2} } B^{\alpha} \Big\} \frac{\sigma^{\alpha} \alpha}{\cos\frac{\pi \alpha}{2} } } \nonumber\\
	= & \lim_{B \rightarrow \infty} \frac{(1-\alpha)  B^{1-2\alpha} }{\exp \Big\{ \frac{\sigma^{\alpha} \alpha}{\cos\frac{\pi \alpha}{2} } B^{\alpha} \Big\} \big(\frac{\sigma^{\alpha}}{\cos\frac{\pi \alpha}{2} }\big)^2 } \nonumber\\
	=& \lim_{B \rightarrow \infty} \frac{(1-\alpha) \cdots (1-(N-2)\alpha)  B^{1-(N-1)\alpha}   }{ \exp \Big\{ \frac{\sigma^{\alpha}}{\cos\frac{\pi \alpha}{2} } B^{\alpha} \Big\} \big(\frac{\sigma^{\alpha} \alpha}{\cos\frac{\pi \alpha}{2} }\big)^{N-1} } \\
	\stackrel{(b)}{=} & \lim_{B \rightarrow \infty} \frac{(1-\alpha) \cdots (1-(N-1)\alpha)  B^{1-N\alpha}   }{ \exp \Big\{ \frac{\sigma^{\alpha}}{\cos\frac{\pi \alpha}{2} } B^{\alpha} \Big\} \big(\frac{\sigma^{\alpha} \alpha}{\cos\frac{\pi \alpha}{2} }\big)^N } \nonumber \\
	=& 0 \nonumber 
\end{align}
where (b) comes from that for a positive $\alpha$, we can always have a positive N so that $1 - (N-1)\alpha >0$ but  $1 - N\alpha <0$.
So we can have the conclusion.
\end{proof}

\bibliographystyle{IEEEtran}
\bibliography{IEEEabrv,draft}

\begin{IEEEbiography}[{\includegraphics[width=1in,height=1.25in,clip,keepaspectratio]{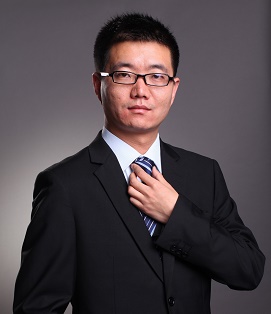}}]{Rongpeng Li}
(S'12-M'17) is now an assistant professor in College of Information Science and Electronic Engineering, Zhejiang University, Hangzhou China. He received his Ph.D and B.E. from Zhejiang University, Hangzhou, China and Xidian University, Xi’an, China in June 2015 and June 2010 respectively, both as “Excellent Graduates”. Dr. Li was a research engineer in Wireless Communication Laboratory, Huawei Technologies Co. Ltd., Shanghai, China from August 2015 to September 2016. He returned to academia in November 2016 as a postdoctoral researcher in College of Computer Science and Technologies, Zhejiang University, Hangzhou, China, which is sponsored by the National Postdoctoral Program for Innovative Talents. His research interests currently focus on Reinforcement Learning, Data Mining and all broad-sense network problems(e.g., resource management, security, etc) and he has authored/coauthored several papers in the related fields. He serves as an Editor of \textsc{China Communications}.\\
\end{IEEEbiography}

\begin{IEEEbiography}[{\includegraphics[width=1in,height=1.25in,clip,keepaspectratio]{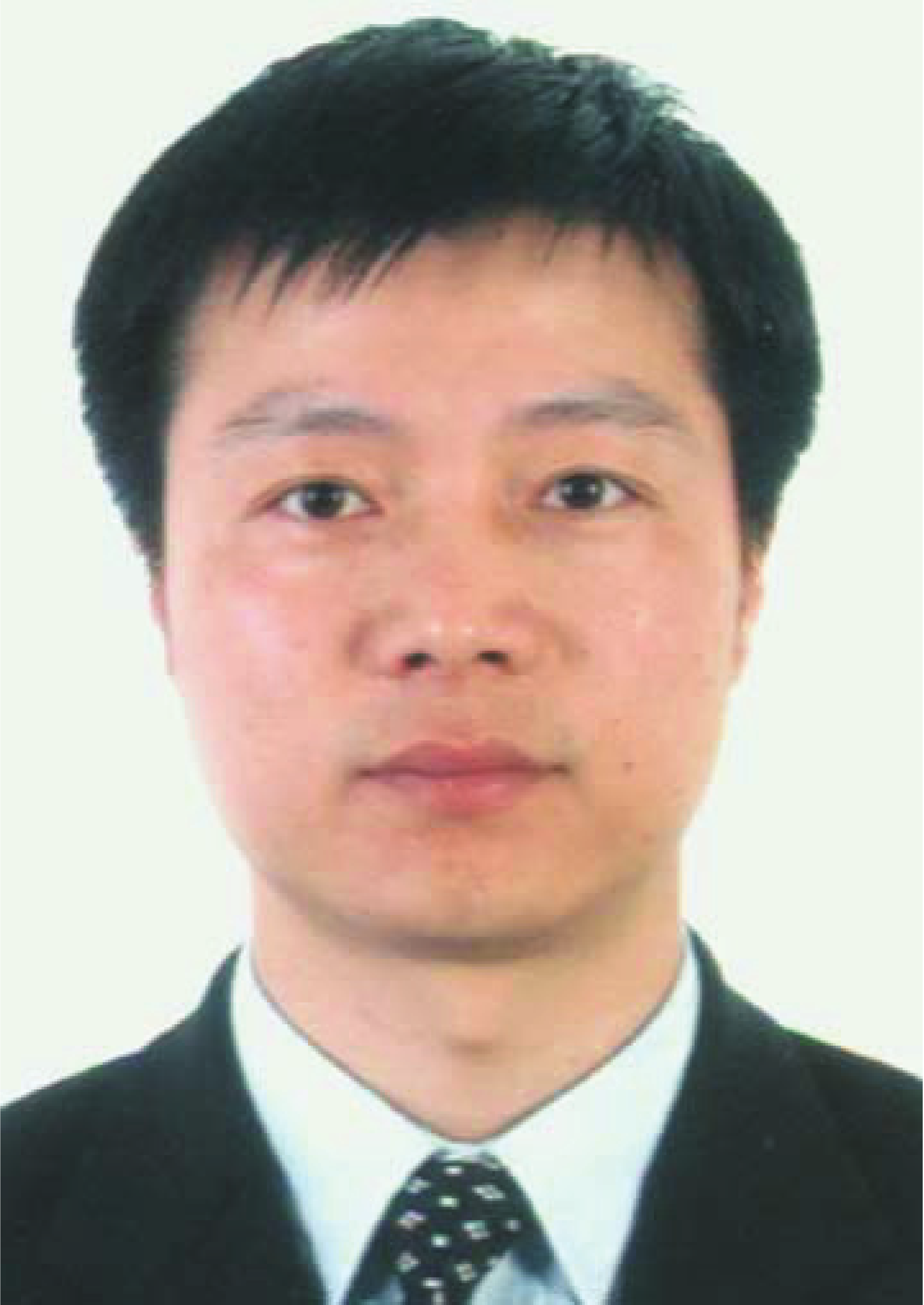}}]{Zhifeng Zhao}
is an Associate Professor at the Department of Information Science and Electronic Engineering, Zhejiang University, China. He received the Ph.D. degree in Communication and Information System from the PLA University of Science and Technology, Nanjing, China, in 2002. Prior to that, he received the Master degree of Communication and Information System in 1999 and Bachelor degree of Computer Science in 1996, from the PLA University of Science and Technology, respectively. From September 2002 to December 2004, he acted as a postdoctoral researcher at the Zhejiang University, where his researches were focused on multimedia NGN (next-generation networks) and soft-switch technology for energy efficiency. From January 2005 to August 2006, he acted as a senior researcher at the PLA University of Science and Technology, Nanjing, China, where he performed research and development on advanced energy-efficient wireless router, Ad Hoc network simulator and cognitive mesh networking test-bed. His research area includes cognitive radio, wireless multi-hop networks (Ad Hoc, Mesh, WSN, etc.), wireless multimedia network and Green Communications. Dr. Zhifeng Zhao is the Symposium Co-Chair of ChinaCom 2009 and 2010. He is the TPC (Technical Program Committee) Co-Chair of IEEE ISCIT 2010 (10th IEEE International Symposium on Communication and Information Technology).
\end{IEEEbiography}

\begin{IEEEbiography}[{\includegraphics[width=1in,height=1.25in,clip,keepaspectratio]{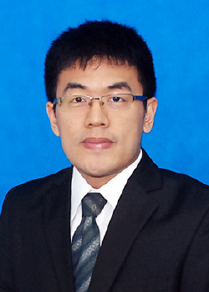}}]{Yi Zhong}
(S'11-M'15) received his B.S. and Ph.D. degree in Electronic Engineering from University of Science and Technology of China (USTC) in 2010 and 2015 respectively. From August to December 2012, he was a visiting student in Prof. Martin Haenggi’s group at University of Notre Dame. From July to October 2013, he was an research intern with Qualcomm Incorporated, Corporate Research and Development, Beijing. From July 2015 to December 2016, he was a Postdoctoral Research Fellow with the Singapore University of Technology and Design (SUTD) in the Wireless Networks and Decision Systems (WNDS) Group led by Prof. Tony Q.S. Quek. Now, he is an assistant professor with School of Electronic Information and Communications, Huazhong University of Science and Technology, Wuhan, China. His main research interests include heterogeneous and femtocell-overlaid cellular networks, wireless ad hoc networks, stochastic geometry and point process theory.
\end{IEEEbiography}

\begin{IEEEbiography}[{\includegraphics[width=1in,height=1.25in,clip,keepaspectratio]{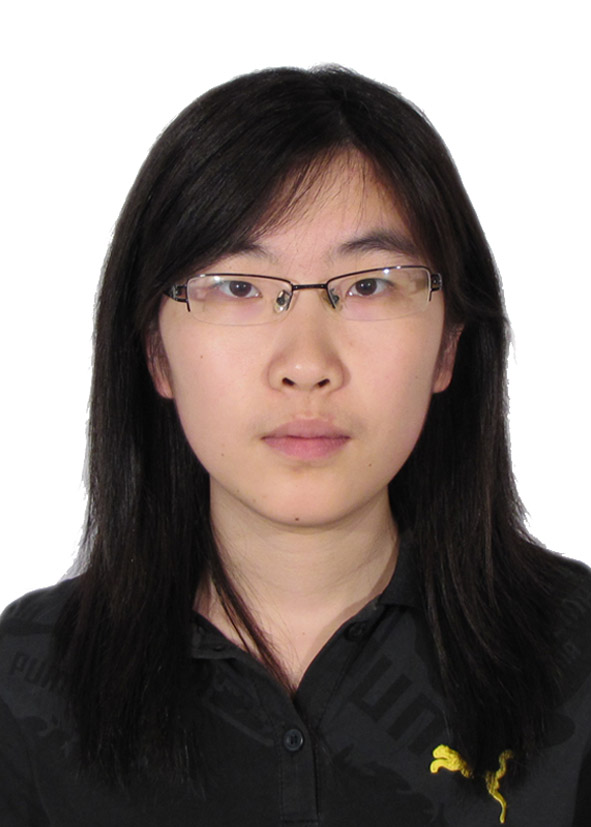}}]{Qi Chen}
received her B.S.in communication engineering from Zhejiang University in 2014. Now, she is a Ph.D. candidate in the College of Information Science and Electrical Engineering, Zhejiang University, Hangzhou, China. Her research interests include data mining of wireless networks, and green communications.
\end{IEEEbiography}

\begin{IEEEbiography}[{\includegraphics[width=1in,height=1.25in,clip,keepaspectratio]{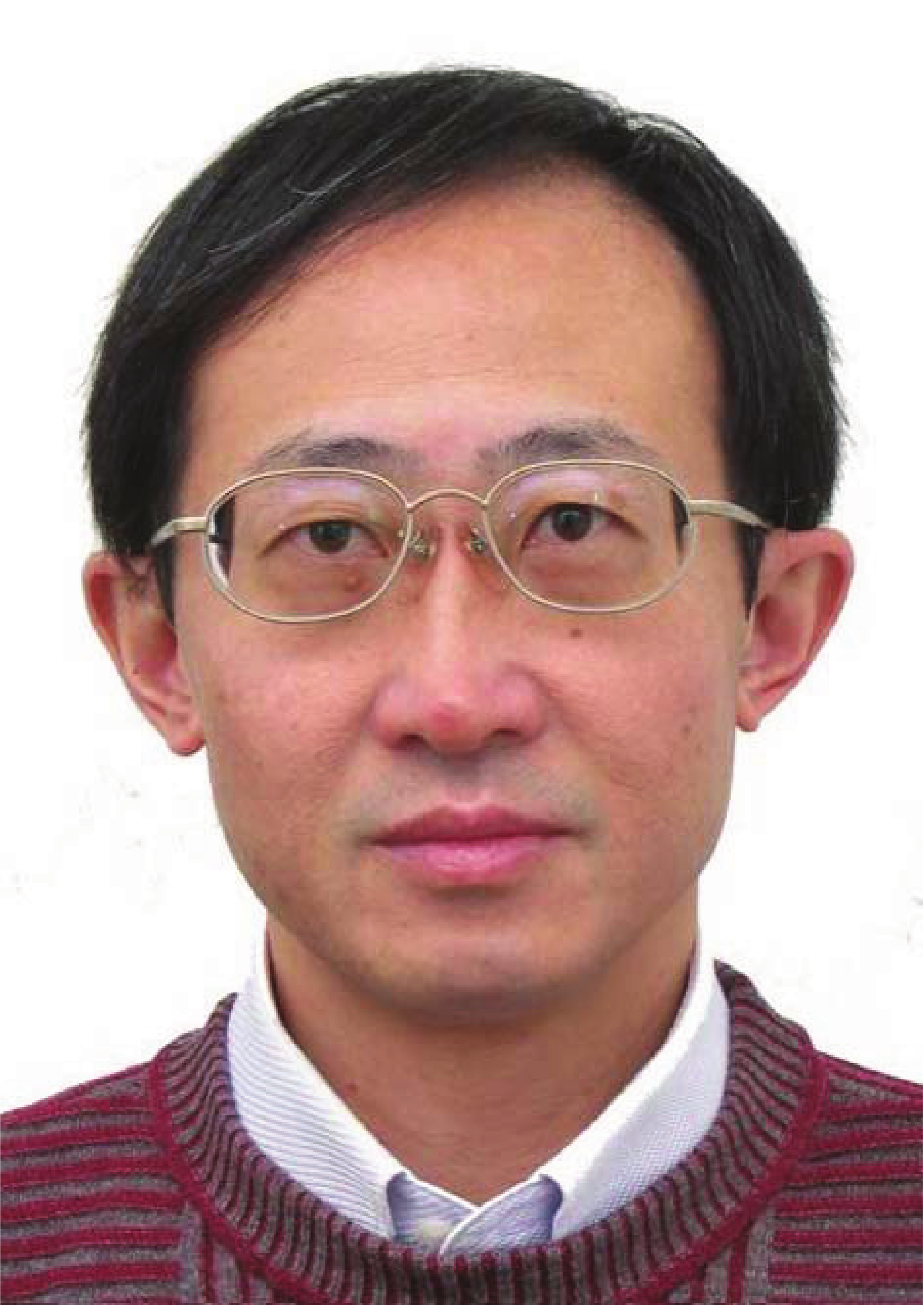}}]{Honggang Zhang}
is a Full Professor with the College of Information Science and Electronic Engineering, Zhejiang University, Hangzhou,  China. He was an Honorary Visiting Professor at the University of York, York, U.K, and the	International Chair Professor of Excellence for Universit\'e Europ\'eenne de Bretagne (UEB) and Sup\`elec, France (2012-2014). He is currently active in the research on	green communications and was the leading Guest Editor of the IEEE Communications Magazine special issues on ``Green Communications". He is taking the role of Associate Editor-in-Chief (AEiC) of China Communications as well as the Series Editors of IEEE Communications Magazine for its Green Communications and Computing Networks Series. He served as the Chair of the Technical Committee on Cognitive Networks of the IEEE Communications Society from 2011 to 2012. He was the co-author and an Editor of two books with the titles of Cognitive Communications-Distributed Artificial Intelligence (DAI), Regulatory Policy and Economics, Implementation (John Wiley \& Sons) and Green Communications: Theoretical Fundamentals, Algorithms and Applications (CRC Press), respectively.
\end{IEEEbiography}

\end{document}